\documentclass[a4paper,11pt]{article}
\pdfoutput=1 
\usepackage{jheppub} 
\usepackage{hyperref}
\usepackage{booktabs}
\usepackage[T1]{fontenc} 
\usepackage{siunitx} 
\usepackage{xcolor,graphicx}
\usepackage{dcolumn}
\usepackage{bm}
\usepackage[normalem]{ulem}
\usepackage{braket} 
\usepackage{amsmath}
\usepackage{cancel}
\usepackage{slashed}
\usepackage{multicol}
\usepackage[capitalise]{cleveref}

\counterwithin*{equation}{section}
\allowdisplaybreaks

\newcommand{\nnl}{\nonumber \\}

\newcommand{\ba}{\begin{array}}
\newcommand{\ea}{\end{array}}
\newcommand{\be}{\begin{eqnarray}}
\newcommand{\ee}{\end{eqnarray} }
\newcommand{\bal}{\begin{align}}
\newcommand{\eal}{\end{align}}
\newcommand{\bi}{\begin{itemize}}
\newcommand{\ei}{\end{itemize}}
\newcommand{\ben}{\begin{enumerate}}
\newcommand{\een}{\end{enumerate}}
\newcommand{\bc}{\begin{center}}
\newcommand{\ec}{\end{center}}
\newcommand{\bt}{\begin{table}}
\newcommand{\et}{\end{table}}
\newcommand{\btb}{\begin{tabular}}
\newcommand{\etb}{\end{tabular}}

\newcommand{\cO}{\mathcal{O}}

\newcommand{\cM}{{\mathcal M}}

\newcommand{\re}{{\mathrm{Re}} \,}
\newcommand{\im}{{\mathrm{Im}} \,}

\newcommand{\hc}{\rm h.c.}

\usepackage{etoolbox,siunitx}
\robustify\bfseries

\begin{document}


\title{On the sensitivity of the $D$ parameter \\  to new physics}

\author[a]{Adam Falkowski,}
\author[a,b]{Antonio Rodr\'iguez-S\'anchez}

\affiliation[a]{Universit\'{e} Paris-Saclay, CNRS/IN2P3, IJCLab, 91405 Orsay, France}

\affiliation[b]{\href{http://www.lpthe.jussieu.fr/spip/index.php}
{\color{black} Sorbonne Universit\'e, CNRS, Laboratoire de Physique Th\'eorique et Hautes Energies (LPTHE), F-75252 Paris, France}}

\emailAdd{adam.falkowski@ijclab.in2p3.fr,
          arodriguez@ijclab.in2p3.fr
          }

\abstract{
Measurements of angular correlations in nuclear beta decay are important tests of the Standard Model (SM). Among those, the so-called $D$ correlation parameter occupies a particular place because it is odd under time reversal, and because the experimental sensitivity is at the $10^{-4}$ level, with plans of further improvement in the near future. Using effective field theory~(EFT) techniques, we reassess its potential to discover or constrain new physics beyond the SM. We provide a comprehensive classification of CP-violating EFT scenarios which generate a shift of the $D$ parameter away from the SM prediction. We show that, in each scenario, a shift larger than $10^{-5}$ is in serious tension with the existing experimental data, where bounds coming from electric dipole moments and LHC observables play a decisive role. The tension can only be avoided by fine tuning of the parameters in the UV completion of the EFT. We illustrate this using examples of leptoquark UV completions. Finally, we comment on the possibility to probe CP-conserving new physics via the $D$ parameter. 
}

\maketitle

\section{Introduction} 

CP violation is an essential ingredient in the fundamental theory of particles and interactions.  
It is present in the Standard Model (SM) in the guise of an invariant phase of the Cabibbo-Kobayashi-Maskawa (CKM) matrix. 
Matter-antimatter asymmetry in the universe strongly hints at the existence of additional sources of CP violation from beyond the Standard Model (BSM). 
On the theory side, our experience with quantum field theory so far suggests that  CP-violating phases in the interaction Lagrangian are generic. 
If that is also the case in the theory underlying the SM, then we expect to eventually observe deviations from the SM predictions in a host of CP-violating observables.  
In fact, several CP-violating observables, such as e.g. the electric dipole moment (EDM) of the neutron or the kaon mass mixing, are potentially sensitive to new physics at enormously high scales, orders of magnitude beyond the direct reach of the Large Hadron Collider (LHC).    
This fact makes such CP-violating observables a likely place where new physics will be {\em first} discovered. 
Conversely, non-observation of non-standard source of CP violation so far provides stringent constraints on virtually every BSM scenario. 
 
In this paper our focus is on CP violation in nuclear beta decay. 
Consider the process ${\cal N} \to {\cal N'} e^- \bar \nu_e$ ($\beta^-$ decay) or  ${\cal N} \to {\cal N'} e^+ \nu_e$ ($\beta^+$ decay), 
where ${\cal N}$ and ${\cal N}'$ the parent and daughter nuclei. 
At the leading (zero-th) order in expansion in $1/m_{\cal N}$, after summing over beta particle and daughter nucleus polarizations, the differential distribution of the decay products takes the most general form~\cite{Jackson:1957zz}
\begin{align}
\label{eq:BETA_dGammaTemplate}
{ d \Gamma \over  d E_e d \Omega_e   d \Omega_\nu  } \sim & 
 1  + b {m_e \over E_e}  
+  a { \boldsymbol{k}_e \cdot \boldsymbol{k}_\nu \over E_e E_\nu} 
 + A {\boldsymbol{J}  \cdot \boldsymbol{k}_e \over J E_e }  
+ B {\boldsymbol{J}  \cdot \boldsymbol{k}_\nu \over J E_\nu }  
\nnl + & 
 \hat c 
 {J(J+1) - 3 (\boldsymbol{J}  \cdot \boldsymbol{j})^2 \over J (J+1)} {(\boldsymbol{k}_e \cdot \boldsymbol{k}_\nu) -    3 (\boldsymbol{k}_e \cdot \boldsymbol{j}) (\boldsymbol{k}_\nu \cdot \boldsymbol{j}) \over 3 E_e E_\nu }
+ {\color{red} D}   
{\boldsymbol{J}  \cdot  (\boldsymbol{k}_e \times \boldsymbol{k}_\nu) \over J E_e E_\nu} , 
\end{align}
where $m_e$ is the electron mass,  
$\boldsymbol{J}$ is the polarization vector of the parent nucleus and $J$ is its spin, 
$\boldsymbol{j}$ is the unit vector in the polarization direction, 
and $\boldsymbol{k}_e$, $\boldsymbol{k}_\nu$, $E_e$, $E_\nu$ are the 3-momenta and the energies of the beta particle and of the neutrino.
The correlation coefficients $b$, $a$, $A$, $B$, and $\hat c$ are $T$-even 
and, with the exception of the Fierz term $b$, they receive $\cO(1)$ contributions in the SM at the leading order, that is at $\cO(\alpha_{\mathrm{EM}}^0)$ 
and at $\cO(1/m_N^0)$ in the non-relativistic expansion in the inverse nucleon mass $m_N$. 
The highlighted correlation coefficient $D$ is on the other hand $T$-odd (because both spin and momenta are T-odd), and is zero in the SM at the leading order.  
The leading SM contributions arise from an interference between one-loop Coulomb corrections and the subleading $\cO(1/m_N^1)$ contributions to the amplitude~\cite{Callan:1967zz}.
Due to the double suppression by $\alpha Z_{\cal N'}$ and $m_e/m_N$, 
the SM contribution is predicted to be small, $|D_{\rm SM}| \lesssim 10^{-4}$, 
and in fact has not been observed experimentally yet in any beta transition.
This is just as good, as it leaves a lot of room to spot non-standard contributions to $D$. 
In particular, CP violation in the fundamental theory underlying the SM could leave an imprint in the form of complex phases of Wilson coefficients in the EFT for beta decay.
Such complex phases would contribute to $D$ at the leading order, potentially inducing $D_{\rm BSM}$ of comparable magnitude to $|D_{\rm SM}|$.  

The current experimental situation regarding the $D$ parameter is summarized in \cref{tab:EFT_D}. 
So far this correlation coefficient was measured in neutron and ${}^{19}$Ne beta decay with the uncertainty of order $10^{-4}$. 
The experimental sensitivity is going to be improved in the near future~\cite{Delahaye:2018kwf}.  
The ongoing experiment MORA at JYFL
will provide a proof-of-principle measurement in ${}^{23}$Mg decay at the $5 \times 10^{-4}$ level. 
Subsequently, measurements at the DESIR facility at GANIL are expected to improve the sensitivity to an $4 \times 10^{-5})$ level. 
Even better sensitivity should be achieved for ${}^{39}$Ca decay, provided a beam with a large enough yield can be produced.

Motivated by this imminent progress,
in this paper we reassess the potential of $D$ measurements to discover or constrain physics beyond the SM~\cite{Herczeg:2001vk,Ng:2011ui,El-Menoufi:2016cfo,Ramsey-Musolf:2020ndm}. 
In \cref{sec:EFT} we present a model-independent analysis employing techniques of effective field theory~(EFT). 
From the low-energy perspective, the $D$ parameter probes certain combinations of the Wilson coefficients in the EFT at the nucleon scale. 
At this level, $D$ provides unique information that is currently unavailable from other probes, in particular about imaginary parts of the Wilson coefficients.  
Furthermore, we connect the nucleon level EFT to more fundamental  EFTs at higher energies, below and above the electroweak scale. 
The latter EFTs are commonly employed in the particle physics literature, and the mapping between their operators and many specific BSM models is well known.  
We classify the CP-violating EFT scenarios according to which operator is responsible for generating the shift  $\Delta D$ of the $D$ parameter away from the SM prediction.   
Then we show that, in each scenario, $|\Delta D| \gtrsim 10^{-5}$ is in serious tension with the existing experimental data, most often with EDMs, but sometimes also with pion decay and/or LHC searches. 
This tension can only be avoided by fine tuning of the parameters in the UV completion of the EFTs.  
In \cref{sec:dr} the discussion is illustrated in concrete BSM settings involving leptoquarks. 
We demonstrate how each EFT scenario leading to  $\Delta D$ can be realized by integrating out leptoquarks with CP-violating couplings to the first generation of the SM fermions.    
Then we take into account the constraints on such leptoquark models  from a host of  low- and high-energy experimental probes, including the direct and indirect searches at the LHC. 
We determine the maximal value of $|\Delta D|$ allowed by existing experimental constraints without fine-tuning. 
The results confirm the earlier EFT estimates, for ${}^{23}$Mg we find $|\Delta D| \lesssim 8 \times 10^{-6}$ in the best case scenario. 

In \cref{sec:cpcon} we try another approach. 
We point out that the $D$ parameter can also probe completely {\em CP-conserving} BSM scenarios. 
In the presence of non-standard scalar and tensor currents in the nucleon-level EFT, leading order contributions to beta decay interfere with the electromagnetic Coulomb corrections. 
This effect contributes to the $D$ parameter even when all the EFT Wilson coefficients are real~\cite{Jackson:1957auh}. 
From this perspective, the $D$ parameter becomes another precision probe of CP-conserving scalar and tensor currents,  on par with more familiar probes in  superallowed $0^+ \to 0^+$, neutron, and mirror beta decays (see~\cite{Gonzalez-Alonso:2018omy} for a review). 
At the moment, the constraints extracted from  $D$ are sill inferior, 
however improving  the sensitivity to the $\cO(10^{-5})$ level will allow one to improve the existing per-mille level bounds obtained in the global analysis in Ref.~\cite{Falkowski:2020pma}. 
 
 \begin{table}[tb]
\bc
\begin{tabular}{|c|c|c|c|c|c|} 
\hline
Parent & $J$ & $r$ & $\kappa_D$ &  $D_{\rm exp}$ & $\Delta D_{\rm future}$ \\ \hline
n & 1/2  & $\sqrt 3$ &  0.88 & $-1.2(2.0)\times 10^{-4}$~\cite{Zyla:2020zbs} & - \\  \hline 
${}^{19}$Ne & 1/2 &  $-1.26$ & $-1.04$ & 0.0001(6)  & - \\  \hline 
${}^{23}$Mg  & 3/2 & -0.44 &  $-1.30$ &  -  & $3.8 \times 10^{-5}$~\cite{Delahaye:2018kwf}   
\\  \hline 
${}^{39}$Ca & 3/2 & 0.52 & $1.42$  &  - & $ 10^{-5}$~\cite{Pierre}  \\  \hline 
 \end{tabular}
\ec
\caption{
\label{tab:EFT_D}
The current experimental measurements and future experimental sensitivity for the D-parameter for various beta transitions.   
We also show the central values of the proportionality constant $\kappa_D$ in the theoretical relation  in \cref{eq:EFT_Dquark} (the errors are small and are not relevant for the present study).}
\end{table}

\section{EFT analysis} 
\label{sec:EFT}

\subsection{Nucleon-level EFT} 

Beta transitions can be described~\cite{Falkowski:2021vdg} in the framework of the pionless EFT~\cite{vanKolck:1999mw}.
The Lagrangian is organized in a non-relativistic expansion in $\boldsymbol{\nabla}/m_N$: \begin{equation}
\label{eq:EFT_pionless}
{\cal L}_{\pi \!\!\! / \rm EFT} \supset 
{\cal L}^{(0)} + {\cal L}^{(1)}  + \cO( \boldsymbol{\nabla}^2/m_N^2) + \hc \, ,
\end{equation} 
where $\boldsymbol{\nabla}$ denotes spatial derivatives and ${\cal L}^{(n)}$ refers to $\cO(\boldsymbol{\nabla}^n/m_N^n)$ terms.
We will focus on the leading order term ${\cal L}^{(0)}$. 
It contains the following interactions relevant for beta decay:\footnote{%
For $A>1$ nuclei one may expect further corrections from two-body currents. However the anatomy of direct matching of quark-level EFT nuclear form factors together with CVC and PCAC relations indicate that the presented picture and the arising conclusions would not become significantly altered by including these corrections.}
\begin{align}
\label{eq:EFT_pionlessL0}
 {\cal L}^{(0)}   \supset & 
 - (\psi_p^\dagger \psi_n)   \bigg  [    
C_V^+   \bar e \bar \sigma^{0} \nu  + C_V^-  e^c \sigma^{0} \bar \nu^c   
+  C_S^+  e^c  \nu   + C_S^- \bar e  \bar \nu^c  
  \bigg ] 
  \nnl &  +  (\psi_p^\dagger \sigma^k \psi_n)   \bigg  [    
  C_A^+ \bar e \bar \sigma^{k}  \nu  +   C_A^- e^c \sigma^{k} \bar \nu^c 
  + C_T^+ e^c \sigma^0 \bar \sigma^{k}  \nu +   C_T^- \bar e \bar \sigma^k \sigma^0  \bar \nu^c \bigg ] ~ . 
\end{align}
The nucleon degrees of freedom are described by non-relativistic quantum fields $\psi_{N, a}$, $N=p,n$, $a=1,2$.  
We work in the isospin limit where the proton and the neutron have the common mass $m_N$. 
For the lepton fields we use the relativistic 2-component spinor notation, following the conventions of Ref.~\cite{Dreiner:2008tw}. 
The sigma matrices are defined as 
$\sigma^\mu = (1, \boldsymbol{\sigma})$, 
$ \bar \sigma^\mu = (1, - \boldsymbol{\sigma})$,
and $\boldsymbol{\sigma} = (\sigma^1,\sigma^2,\sigma^3)$ is a 3-vector of  the usual Pauli matrices.   
In this language,  $\psi$ and $\bar \psi^c$ correspond to the left- and right-handed components of a spin-1/2 Dirac fermion. 
The connection to the 4-component notation in the chiral representation is 
$\Psi = \begin{pmatrix}  \psi_\alpha \\ \bar \psi^c{}^{\dot \beta}  \end{pmatrix}$, 
$\bar \Psi = \begin{pmatrix}   \psi^c{}^{\alpha}  \\ \bar \psi_{\dot \beta}  \end{pmatrix}$, 
$\gamma^\mu = \begin{pmatrix} 0 & \sigma^\mu \\ \bar \sigma^\mu & 0  \end{pmatrix}$. 
To ease the comparison with other works let us note that the corresponding identities involving Lorentz tensor structures made out of fermion bilinears can be trivially recovered from
\begin{align}\nonumber
\bar{\psi}_1\bar{\psi}_2^{c}&=\bar{\Psi}_{1L}\Psi_{2R}  &  \bar{\psi}_{1}\bar{\sigma}^{\mu}\psi_2&=\bar{\Psi}_{1L}\gamma^{\mu}\Psi_{2L}&  \bar{\psi}_{1}\bar{\sigma}^{\mu\nu}\bar{\psi}_2^c&=\bar{\Psi}_{1L}\Sigma^{\mu\nu}\Psi_{2R} \, ,
\\
\psi_1^c\psi_2&=\bar{\Psi}_{1R}\Psi_{2L}&  \psi_{1}^c\sigma^{\mu}\bar{\psi}_2^c&=\bar{\Psi}_{1R}\gamma^{\mu}\Psi_{2R}&  \psi_{1}^c\sigma^{\mu\nu}\psi_2&=\bar{\Psi}_{1R}\Sigma^{\mu\nu}\Psi_{2L} \, ,
\end{align}
and taking into account that the same identities hold when simultaneously changing any (barred or not) $\psi_{i}\leftrightarrow \psi_{i}^{c}$ and $\Psi_{iL}\leftrightarrow \Psi_{iR}^{C}$; and/or $\psi_{j}^c\leftrightarrow \psi_{j}$ and $\Psi_{jR}\leftrightarrow \Psi_{jL}^{C}$,
where $\Psi_{L(R)}^{C}\equiv C \bar{\Psi}_{L(R)}^{T}$. We have defined $\Sigma^{\mu\nu}=\frac{i}{2}[\gamma^{\mu},\gamma^{\nu}]$, $\sigma^{\mu\nu}\equiv \frac{i}{2}(\sigma^{\mu}\bar{\sigma}^{\nu}-\sigma^{\nu}\bar{\sigma}^{\mu})$ and $\bar{\sigma}^{\mu\nu}\equiv \frac{i}{2}(\bar{\sigma}^{\mu}\sigma^{\nu}-\bar{\sigma}^{\nu}\sigma^{\mu})$.

The neutrinos are treated as massless and we allow for the possibility of right-handed neutrinos contributing to beta decay.  
For the Wilson coefficients we use the conventions of Ref.~\cite{Falkowski:2020pma}, where $C_X^+$ ($C_X^-$) parametrize interactions of left-handed (right-handed) neutrinos. Our conventions are simply related to the commonly used $C_X$ and $C_X'$ variables introduced by  Lee and Yang~\cite{Lee:1956qn}:
$C_X = (C_X^+ + C_X^-)/2$, $C_X' = (C_X^+ - C_X^-)/2$. 
The situation where the right-handed neutrino is absent from the low-energy EFT (e.g. because it has a large Majorana mass) can be described by setting  $C_X^- = 0$ for all $X$.

For the same spin of the parent and daughter nuclei, $J=J'$, the $D$ correlation defined by \cref{eq:BETA_dGammaTemplate} depends on the Wilson coefficients in \cref{eq:EFT_pionlessL0} as 
\begin{equation}
\label{eq:EFT_Dnucleon}
D  =   - 2 r \sqrt{J \over J+1}  { \im \bigg \{   
C_V^+ \bar C_A^+ - C_S^+ \bar C_T^+
+ C_V^- \bar C_A^-  - C_S^- \bar C_T^-  \bigg \}   
\over 
 |C_V^+|^2   + |C_S^+|^2  +  |C_V^-|^2   + |C_S^-|^2  
+  r^2 \big [  |C_A^+|^2  + |C_T^+|^2  +  |C_A^-|^2  + |C_T^-|^2 \big ] },  
\end{equation}
Here $r$ is the ratio of the Gamow-Teller and Fermi matrix elements. 
For the neutron decay $r=\sqrt 3$, while for nuclei with the mass number $A>1$ it can be extracted from experimental data. 
Note that the D parameter is non-zero only if at least some of the Wilson coefficients have distinct complex phases. 
 
\subsection{Quark-level EFT}

At a more fundamental level, nuclear beta decays probe charged-current interactions between the first generation of quarks and leptons.
We consider an EFT for these degrees of freedom valid between the scales of $\sim 2$~GeV and the electroweak scale $\sim m_W$.
We will refer to this EFT as the $\nu$WEFT.  
The leading order effective interactions contributing to beta decay can be described by the following relativistic Lagrangian:
\begin{align}
\label{eq:EFT_Lrweft}
{\cal L}_{\nu \rm WEFT}  \supset 
- \frac{2 V_{ud}}{v^2} \Big \{ 
& \left( 1 +  \epsilon_L \right) \
 (\bar{e}  \bar \sigma_\mu  \nu)(\bar u \bar \sigma^\mu d  )
~+~
\tilde \epsilon_L  (e^c  \sigma_\mu \bar \nu^c) (\bar u \bar \sigma^\mu d)
\nnl
 + &  \epsilon_R  (\bar{e}  \bar \sigma_\mu  \nu) (u^c   \sigma^\mu \bar d^c)
~+~
 \tilde \epsilon_R  (e^c  \sigma_\mu \bar \nu^c)(u^c   \sigma^\mu \bar d^c)
 \nnl
 +  & {1 \over 2} (e^c  \nu)  \big [   (\epsilon_S + \epsilon_P ) u^c d  + (\epsilon_S- \epsilon_P ) \bar u \bar d^c     \big ] 
~+~ {1 \over 2} (\bar{e}  \bar \nu^c ) \big [  
(\tilde \epsilon_S + \tilde \epsilon_P ) u^c d  + (\tilde \epsilon_S-  \tilde \epsilon_P ) \bar u \bar d^c    \big ] 
\nnl
+ & {1 \over 4}  \epsilon_T  (e^c   \sigma_{\mu \nu} \nu) (u^c   \sigma^{\mu \nu} d)
~+~
{1 \over 4}  \tilde \epsilon_T (\bar{e}   \bar \sigma_{\mu \nu}  \bar \nu^c)(\bar u  \bar \sigma^{\mu \nu} \bar d^c)
\Big \}  + {\rm h.c.}~ \qquad 
\end{align}
where $u$ and  $d$ are the up quark, down quark, 
$V_{ud}$ is an element of the unitary CKM matrix, 
and $v \approx 246.22$~GeV is related by $G_F = (\sqrt 2 v^2)^{-1}$ to the Fermi constant $G_F$ measured in muon decay. 
 The Wilson coefficients $\epsilon_X$ and $\tilde \epsilon_X$, 
$X = L,R,S,P,T$, parametrize non-SM  effects, and $\epsilon_X = \tilde \epsilon_X = 0$ in the SM limit.

At tree level, the map between the Wilson coefficients in \cref{eq:EFT_pionlessL0} and in \cref{eq:EFT_Lrweft} is given by~\cite{Gonzalez-Alonso:2018omy}
\begin{align}
\label{eq:EFT_mapCtoEpsilon}
C_V^+  = & {V_{ud} \over  v^2} g_V \big ( 1+ \epsilon_L + \epsilon_R \big ) , 
\qquad 
C_V^-  =  {V_{ud} \over v^2} g_V  \big ( \tilde \epsilon_L + \tilde \epsilon_R \big ) , 
\nnl 
C_A^+ = & - {V_{ud} \over v^2} g_A \big ( 1+ \epsilon_L - \epsilon_R \big ) , 
\qquad 
C_A^-  =  {V_{ud} \over v^2} g_A  \big ( \tilde \epsilon_L - \tilde \epsilon_R \big ) , 
\nnl
C_T^+  = & {V_{ud} \over v^2} g_T \epsilon_T  , 
\qquad 
C_T^-  =  {V_{ud} \over v^2} g_T \tilde \epsilon_T, 
\nnl
C_S^+  = & {V_{ud} \over v^2} g_S \epsilon_S  , 
\qquad 
C_S^-  =  {V_{ud} \over  v^2} g_S \tilde \epsilon_S. 
\end{align}
Here, $g_{V,A,S,T}$ are  non-perturbative parameters referred to as the vector, axial, scalar, and tensor charges of the nucleon. 
For the vector charge,  $g_V = 1$ up to (negligible) quadratic corrections in isospin-symmetry breaking~\cite{Ademollo:1964sr}. 
The remaining charges are not known from symmetry considerations alone and must be fixed from experimental data or by lattice calculations. 
In this work we will use the FLAG'21 values: 
$g_A = 1.246(28)$~\cite{Aoki:2021kgd,Gupta:2018qil,Chang:2018uxx,Walker-Loud:2019cif},  $g_S = 1.022(100)$, and $g_T = 0.989(34)$~\cite{Aoki:2021kgd,Gupta:2018qil}. 

Using this map, we can translate \cref{eq:EFT_Dnucleon} into the quark-level Wilson coefficients:
\begin{equation}
D  \approx  {4 r g_V g_A \over g_V^2 + r^2 g_A^2} \sqrt{J \over J+1}  \im 
\bigg[ \epsilon_R(1 + \epsilon_L^*) + {g_S g_T \over 2 g_V g_A} (\epsilon_S   \epsilon_T^* +  \tilde \epsilon_S  \tilde \epsilon_T^* )   
 - \tilde \epsilon_R  \tilde \epsilon_L^*   \bigg ],   
\end{equation} 
where we neglected the new physics corrections originating from the denominator of \cref{eq:EFT_Dnucleon}, but we kept the linear and quadratic effects in Wilson coefficients coming from the numerator. 
We can recast the above in the semi-numerical form as 

\begin{equation}
\label{eq:EFT_Dquark}
D    \approx  \kappa_D \,    \im  \big [  \epsilon_R(1 + \epsilon_L^*)  
 + 0.4  (\epsilon_S   \epsilon_T^* +  \tilde \epsilon_S  \tilde \epsilon_T^* ) 
  - \tilde \epsilon_R  \tilde \epsilon_L^* ] , \qquad 
 \kappa_D \equiv  {4 r g_V g_A \over g_V^2 + r^2 g_A^2} \sqrt{J \over J+1} . 
\end{equation} 
This is the master equation for the D parameter that we will use extensively in the following.  
The values of the proportionality constant $\kappa_D$ for selected beta transitions are displayed in \cref{tab:EFT_D}. 
At the linear level, the D parameter only probes CP violation  entering via the so-called  right-handed currents, that is the effective weak interactions between left-handed leptons and right-handed quarks. 
At the quadratic level, other non-standard currents are probed as well, in particular the scalar and tensor currents involving the left- and right-handed neutrinos.  

\subsection{EFT above the electroweak scale}

We move to discussing the effective theory above the electroweak scale that, under very broad assumptions,  UV-completes the $\nu$WEFT, which is often referred to  as the $\nu$SMEFT~\cite{Liao:2016qyd,Li:2021tsq}. 
It has the gauge symmetry $SU(3)\times SU(2)\times U(1)$ and the degrees of freedom are those of the SM plus three generations of  right-handed neutrinos which are gauge singlets. 
The Lagrangian consists of all independent gauge invariant operators made of these fields organized in the expansion in powers of $1/\Lambda$, according to the canonical dimensions of the operators. 
We will discuss dimension-6 and dimension-8 operators that can induce the (tilde)  $\epsilon_X$ Wilson coefficients in the Lagrangian \cref{eq:EFT_Lrweft} below the electroweak scale. 
We will be interested in generating $\epsilon_X$ entering the master equation (\ref{eq:EFT_Dquark}), leading to a non-zero value of the D-parameter. 
The master equation contains four distinct contributions in the square bracket, and we will discuss in turn how to generate the corresponding Wilson coefficients.  

\vspace{1cm}

{\bf Scenario I}. 
We start with operators generating $\epsilon_R$ below the electroweak scale.   
One possible source is the dimension-6 operator 
\begin{equation}
\label{eq:EFT_scenario1a}
{\cal L }_{\nu \rm SMEFT}  \supset  i C_{\phi u d}  \tilde H^\dagger D_\mu H (u^c \sigma^\mu \bar d^c)  + \hc , 
\end{equation}
where $H$ is the SM Higgs doublet field, and $\tilde H_a = \epsilon^{ab} H^*_b$ (in our conventions the Higgs VEV is given by $\langle H^T\rangle  = (0,v)/\sqrt 2$ with   
$v\approx 246\, \mathrm{GeV}$).
This operator induces a coupling of the $W$ boson to the right-handed up and down quarks: 
\begin{equation}
{\cal L}_{\nu \rm SMEFT}  \supset  {g_L \over \sqrt 2} W_\mu^+   \left [ \bar \nu \bar \sigma^\mu  e +  V_{ud}  \bar u \bar \sigma^\mu d 
+ {v^2 \over 2} C_{\phi u d}   u^c  \sigma^\mu \bar d^c  \right ]  + \hc  
\end{equation}
Integrating out the $W$ boson at tree level, below the electroweak scale one finds the 4-fermion interaction $(\bar e  \bar \sigma_\mu \nu) (u^c   \sigma^\mu   \bar d^c)$ from \cref{eq:EFT_Lrweft} with the Wilson coefficient
\begin{equation}
\epsilon_R =  {v^2 \over 2 V_{ud}} C_{\phi u d} .
\end{equation}
Consequently, in this scenario the D parameter depends on the $\nu$SMEFT Wilson coefficients as
\begin{equation}
\label{eq:EFT_Dscenario1a}
D    \approx  {\kappa_D  \over 2}   \im  \big [  v^2 C_{\phi u d}  \big ]   \, . 
\end{equation}
Therefore one way to induce the D parameter is to generate the operators in \cref{eq:EFT_scenario1a} with a complex Wilson coefficient $C_{\phi u d}$ in the EFT above the electroweak scale. 
This option may seem promising because the D-parameter appears at $\cO(\Lambda^{-2})$, and then it can be sizable even if the BSM scale $\Lambda$ is relatively large. Moreover, the Wilson coefficient $C_{\phi u d}$ is induced by several well-motivated BSM models, in particular by the left-right symmetric models~\cite{Pati:1974yy}. 
However,  this scenario faces one phenomenological problem~\cite{Ng:2011ui}, which can already be seen at the EFT level.  
The problem stems from the fact that, together with $\epsilon_R$, another 4-fermion operator is generated below the electroweak scale: 
\begin{equation}
\label{eq:EFT_C1LR}
{\cal L}_{\nu \rm SMEFT} \supset 
-  C_{1LR}  (\bar d \bar \sigma_\mu u )  (u^c \sigma^\mu \bar d^c) + \hc 
\end{equation}
with  $C_{1LR} = V_{ud} C_{\phi u d}$.  
Thus the magnitude and phase of $C_{1LR}$ is perfectly correlated with that of $\epsilon_R$, which is at the origin of the $D$ parameter. 
On the other hand, the imaginary part of  $C_{1LR}$ is strongly constrained  by nuclear EDMs. 
Formally the strongest constraint comes from the measurement of EDM in mercury, but the relation between  $d_{{}^{199}Hg}$ and  ${\rm Im} C_{1LR}$ suffers from large theoretical uncertainties.
To be conservative, here we will use the slightly weaker but theoretically more robust constraint from the neutron EDM measurement.
Using~\cite{Alioli:2017ces} 
\begin{equation}
\label{eq:EFT_dneutron}
d_n  =   (22 \pm 14)\, v^2  \,{\rm Im} C_{1LR} \times 10^{-22} e \, {\rm cm} 
\end{equation}
and the current best measurement  $d_n = (0.0 \pm 1.1) \times 10^{-26} e \, {\rm cm} $~\cite{Abel:2020pzs} we get the following $95\%$~CL constraint\footnote{%
Using the information from nuclear EDMs Ref.~\cite{Ramsey-Musolf:2020ndm} quotes a slightly stronger constraint $v^2 |{\rm Im} C_{1LR}| \leq 3 \times 10^{-6}$. 
} at $\mu=2\, \mathrm{GeV}$
\begin{equation}
v^2 |{\rm Im} C_{1LR}| \lesssim 1 \times 10^{-5} . 
\end{equation}
All in all, in this scenario  we can relate the D parameter to the strongly constrained imaginary part of $C_{1LR}$:  
\begin{equation}
|D|    \approx  5  \times 10^{-6}  {v^2 |{\rm Im} C_{1LR}|  \over 10^{-5}} \times \kappa_D . 
\end{equation}  
This implies that $|D| < 10^{-5}$  generically, which is below the experimental sensitivity in the near future. 
A larger $D$ parameter can be achieved if one allows for some fine-tuning between different contributions to $d_n$. 
For example, one can arrange for a partial cancellation between the contributions proportional to $\im  C_{1LR}$ and those proportional to the QCD $\theta$ parameter~\cite{Ramsey-Musolf:2020ndm}.  

Another option is to generate  $\epsilon_R$ from the dimension-8 operator \begin{equation}
\label{eq:EFT_scenario1b}
{\cal L}_{\nu \rm SMEFT}  \supset  C_8 (\bar l H \bar \sigma_\mu H \tilde l)   (u^c  \sigma^\mu \bar d^c) , 
\end{equation} 
where $l = (\nu,e)$ is the 1st generation lepton doublet, and 
$\tilde l_a \equiv \epsilon^{ab} l_b$.
Once the Higgs field acquires VEV, it generates the $(\bar e  \bar \sigma_\mu \nu) (\bar{u}^c   \sigma^\mu   \bar d^c)$ operator from \cref{eq:EFT_Lrweft} with the Wilson coefficient
\begin{equation}
\epsilon_R =  {v^4 C_8 \over 4 V_{ud}} . 
\end{equation} 
The D parameter depends on the dimension-8 $\nu$SMEFT Wilson coefficient as
\begin{equation}
\label{eq:EFT_Dscenario1b}
D    \approx  {\kappa_D  \over 4}   \im  \big [  v^4 C_8  \big ].    
\end{equation}
The advantage of generating $\epsilon_R$ via the dimension-8 operator in \cref{eq:EFT_scenario1b} is that the 4-quark operator in \cref{eq:EFT_C1LR} is not generated at tree level. 
It is however generated at one loop in the EFT, and it is quadratically divergent. 
The divergence means that the associated contribution to the EDM is not calculable within the EFT.\footnote{Technically within the EFT these contributions are absent  in $\mathrm{MS}$-like schemes, which preserves the dimensional counting beyond tree level. But then one generically expects that the very same UV interaction that induces the $D=8$ contribution at tree level will also induce the problematic $D=6$ one at one-loop level when matching with the EFT.
}
Instead, the result depends on the UV completion.   
Nevertheless, one can estimate
\begin{equation} 
C_{1LR} \sim {\Lambda^2 \over 4 \pi^2} C_8,  
\end{equation}
where $\Lambda$ is the mass scale of the BSM particles that generate the operator in  \cref{eq:EFT_scenario1b} and regularize the quadratic divergence.  
The D parameter can then be estimated as 
\begin{equation}
D \sim  \kappa_D  10^{-4} 
\bigg ( {v^2  \im  [   C_{1LR} ]   \over 10^{-5}} \bigg )  {v^2 \over \Lambda^2 }. 
\end{equation}
A $D$ parameter of order $10^{-4}$ can be obtained, but only when the BSM particles are near the electroweak scale. 
Therefore, in this scenario it is vital to discuss experimental constraints on the possible UV completions, as the new particles may be within the reach of the LHC. 
We will discuss this issue later in this paper in the context of leptoquark UV completions.
As soon as the new particles are far above the TeV scale, the operator in \cref{eq:EFT_Dscenario1b} again leads to the D parameter being suppressed to a currently unobservable level.

\vspace{1cm}

{\bf Scenario II}. 
We move now to the situation where the D parameter is generated via the $\im[\epsilon_S \epsilon_T^*]$ term in \cref{eq:EFT_Dquark}, that is via the scalar and tensor interactions with left-handed neutrinos. 
The following operators in the $\nu$SMEFT induce these Wilson coefficients below the electroweak scale:  
\begin{equation}
\label{eq:EFT_scenario2}
{\cal L}_{\nu \rm SMEFT} \supset 
C_{lequ}^{(1)}  \epsilon^{ab} (\bar l_a \bar e^c)(\bar q_{b} \bar u^c) 
+ C_{lequ}^{(3)}  \epsilon^{ab} (\bar l_a \bar \sigma^{\mu\nu} \bar e^c)(\bar q_{b} \bar \sigma_{\mu\nu}  \bar u^c) 
+C_{ledq} (\bar l_a \bar e^c)(d^c q_a )   
+ \hc
\end{equation} 
Above $q$ is the 1st generation quark doublet field. 
We work in the down-type basis here $q = (V_{ud}^* u + V_{cd}^* c  + V_{td}^* t,d)$. 
Matching this at tree level to the $\nu$WEFT Lagrangian in \cref{eq:EFT_Lrweft} we get 
\begin{align}
\epsilon_S  =  &  -{v^2 \over 2 V_{ud}} 
\bigg [ C_{lequ}^{(1)*} + V_{ud} C_{ledq}^*  \bigg ] , 
\nnl 
\epsilon_P  =  &  -{v^2 \over 2 V_{ud}} 
\bigg [ C_{lequ}^{(1)*} - V_{ud} C_{ledq}^*  \bigg ],  
\nnl 
\epsilon_T  =  &  -{2 v^2 \over  V_{ud}} C_{lequ}^{(3)*} . 
\end{align} 
The contribution to the $D$ parameter is then
\begin{equation}
  \label{eq:EFT_Dscenario2}
D \approx - 0.4  \kappa_D  v^4
 \im \bigg\{ \big [ C_{lequ}^{(1)} + V_{ud} C_{ledq} \big ] C_{lequ}^{(3)*}  \bigg \}  . 
  \end{equation} 
In order to generate the D parameter in this scenario one needs to induce the tensor Wilson coefficient $C_{lequ}^{(3)}$ simultaneously with and with a different phase than $C_{lequ}^{(1)}$ and/or $C_{ledq}$. 
As we will discuss later on, this is possible e.g. in leptoquark models.
However, this scenario faces a disastrous problem already at the EFT level.
 
The point is that the operators in Eq.~(\ref{eq:EFT_scenario2}), in addition to the charged currents contributing to beta decays, also induce the neutral current interactions: 
\begin{equation}
\label{eq:EFT_scenario2NC}
{\cal L}_{\nu \rm WEFT} \supset 
-C_{lequ}^{(1)} V_{ud} (\bar e \bar e^c)   (\bar u \bar u^c) 
- C_{lequ}^{(3)}  V_{ud} (\bar e \bar \sigma^{\mu\nu} \bar e^c)(\bar u \bar \sigma_{\mu\nu}  \bar u^c) 
+C_{ledq} (\bar e \bar e^c)(d^c d )   
+ \hc
\end{equation} 
An imaginary part in any of the Wilson coefficients would induce EDMs in electron, nucleons, nuclei and atoms, which are very strongly constrained by current bounds. 
Using the theoretical expressions from Refs.~\cite{Dekens:2018bci,Dekens:2018pbu,Aebischer:2021uvt},  
the EDM measurement using the ThO molecule~\cite{ACME:2018yjb} we find the constraints 
\begin{equation}
\label{eq:EFT_scenario2EDM}
v^2 |\mathrm{Im}C_{lequ}^{(1)}| \lesssim 1 \times 10^{-10}, 
\quad 
v^2 |\mathrm{Im}C_{lequ}^{(3)}| \lesssim 5 \times 10^{-11}, 
\quad
v^2 |\mathrm{Im}C_{ledq}| \lesssim 1 \times 10^{-10}, 
\end{equation}
at $95\%$~CL. 
This constrains the imaginary parts of $\epsilon_S$ and $\epsilon_T$ to $O(10^{-10})$ level. Since the real parts are constrained at the $C(10^{-3})$ level~\cite{Falkowski:2020pma}, one concludes that in this scenario
\begin{equation}
|D| \lesssim 10^{-13}     . 
\end{equation}
It is safe to state that, in this scenario,  BSM contributions to the D parameter will never be experimentally observed.

\vspace{1cm}

{\bf Scenario III}. 
New limiting factors arise when the $D$ parameter is generated via the scalar and tensor interactions involving right-handed neutrinos, cf. the third term in \cref{eq:EFT_Dquark}. 
The $\nu$SMEFT operators relevant for this scenario are
\begin{equation}
\label{eq:EFT_scenario3}
{\cal L}_{\nu \rm SMEFT} \supset 
C_{l\nu qd}^{(1)}  \epsilon^{ab} (\bar l_a \bar \nu^c)( \bar q_{b} \bar d^c )  
+  C_{l\nu qd}^{(3)}  \epsilon^{ab} (\bar l_a \bar \sigma^{\mu\nu} \bar \nu^c)(\bar q_{b} \bar \sigma_{\mu\nu}  \bar d^c) 
+ C_{l\nu uq} (\bar l_a \bar \nu^c) (u^c q_{a}  )
+ \hc
\end{equation} 
Matching this to the quark-level Lagrangian in \cref{eq:EFT_Lrweft} we get 
\begin{align}
  \label{eq:EFT_scenario3matcing}
\tilde \epsilon_S  =  &  {v^2 \over 2 V_{ud}} \bigg [  
V_{ud}C_{l\nu qd}^{(1)}  -  C_{l\nu uq}   \bigg ] , 
\nnl 
\tilde \epsilon_P  =  &  -{v^2 \over 2 V_{ud}} \bigg [ 
V_{ud} C_{l\nu qd}^{(1)}  +  C_{l\nu uq}   \bigg ],  
\nnl 
\tilde \epsilon_T  =  &  2 v^2  C_{l\nu qd}^{(3)} , 
\end{align}
and the D-parameter expressed by ${\nu \rm SMEFT}$ Wilson coefficients  reads

\begin{equation}
  \label{eq:EFT_Dscenario3}
D \approx 0.4 \kappa_D v^4 
 \im \bigg \{ \big [ 
V_{ud} C_{l\nu qd}^{(1)}  - C_{l\nu uq}  \big ] C_{l\nu qd}^{(3)*} \bigg \}  . 
  \end{equation}
The main problem is  that, as can be seen from \cref{eq:EFT_scenario3matcing}, in this scenario $\tilde \epsilon_P$ is generated together with $\tilde\epsilon_S$ and with the same order of magnitude. 
The former is strongly constrained by pion decay due to the chiral enhancement of the pseudoscalar contribution by the large factor $m_\pi^2/m_e(m_u+m_d)$. 
The measurement 
${\Gamma(\pi \to e \nu) \over \Gamma(\pi \to \mu\nu)} = 1.2327(23) \times 10^{-4}$~\cite{Zyla:2020zbs} translates to 
$|\tilde \epsilon_P| < 1.0 \times 10^{-5}$ at 95\% CL. 
This leads to the suppression of the $D$ parameter by the factor of $\cO(10^{-5})$, unless $\tilde{\epsilon}_S \sim \tilde{\epsilon}_P$ is avoided via fine-tuning or savvy model building.

Another set of constraints arises due to the neutral currents predicted by \cref{eq:EFT_scenario3}: 
\begin{equation}
\label{eq:EFT_scenario3nc}
{\cal L}_{\nu \rm SMEFT} \supset 
C_{l\nu qd}^{(1)}   (\bar \nu \bar \nu^c)( \bar d  \bar d^c )  
+  V_{ud} C_{l\nu uq} (\bar \nu \bar \nu^c) (u^c u)
+ C_{l\nu qd}^{(3)}  (\bar \nu \bar \sigma^{\mu\nu} \bar \nu^c)(\bar d  \bar \sigma_{\mu\nu}  \bar d^c) 
+ \hc
\end{equation} 
The first two terms above contribute to neutrino masses after QCD phase transition.
Naturalness then dictates $\Lambda_{\rm QCD}^3 (C_{l\nu qd}^{(1)} +  V_{ud} C_{l\nu uq}) \lesssim 0.1$~eV, unless there exists some mechanism ensuring cancellation of this contribution against that from the usual Yukawa coupling of the neutrinos. 
This translates to $|\tilde{\epsilon}_S| \lesssim 10^{-3}$.  
The last term induces the neutrino magnetic moment 
$\mu_\nu \sim \Lambda_1 \re C_{l\nu qd}^{(3)}$, where $\Lambda_1 \sim 10$~MeV~\cite{Dekens:2018pbu}. 
Given the experimental constraint $\mu_\nu \lesssim 3 \times 10^{-11} \mu_B$~\cite{Beda:2013mta,Borexino:2017fbd}, where $\mu_B$ is the Bohr magneton,  
one obtains $\re \tilde \epsilon_T \lesssim 0.1$. 
Similar constraint, $|\tilde \epsilon_T| \lesssim 0.1$, can be obtained from the global fit to nuclear beta decay~\cite{Falkowski:2020pma}.

All in all, the $D$ parameter in this scenario  can be written as 
\begin{equation}
D \sim 10^{-6} \kappa_D  \im \bigg [ \bigg ( {\tilde \epsilon_T \over 10^{-1}} \bigg )   
\bigg ( {\tilde \epsilon_S \over 10^{-5} } \bigg)   \bigg ] ,  
\end{equation}
and thus $|D| \lesssim 10^{-6}$ in the absence of fine-tuning. 
Pushing $D$ to the observable level requires a similar level of fine-tuning as in  scenario I where it appears through $\epsilon_R$. 
  
\vspace{1cm}

{\bf Scenario IV}. 
We finish our EFT exploration by discussing a set-up where the D parameter is generated via the last term in \cref{eq:EFT_Dquark} proportional to $\im[\tilde \epsilon_L \tilde \epsilon_R^*]$
Consider the following dimension-6 $\nu$SMEFT operators:
 \begin{equation}
 \label{eq:EFT_scenario4}
 {\cal L}_{\nu \rm SMEFT} \supset   
 C_{e \nu u d} (e^c \sigma^\mu \bar \nu^c)(u^c \sigma_\mu \bar d^c)
 + i C_{\phi e \nu}  D_\mu H^\dagger \tilde H (e^c \sigma^\mu \bar \nu^c) + \hc  
 \end{equation} 
 The first term maps directly to $\tilde \epsilon_R$. 
 The second induces, after electroweak symmetry breaking, the non-standard $W$ boson interactions right-handed leptons: 
 \begin{equation}
 {\cal L}_{\nu \rm SMEFT} \supset 
 - {g v^2 \over 2 \sqrt 2 }   C_{\phi e \nu}^* W_\mu^+  (\nu^c \sigma^\mu \bar e^c) + \hc 
 \end{equation}  
 Integrating out the W boson one obtains 
 \begin{equation}
 {\cal L}_{\nu \rm WEFT} \supset - {g^2 \over 2 m_W^2} 
 \big |   V_{ud}  (\bar u \bar \sigma_\mu d) - {v^2 \over 2} C_{\phi e \nu}^*  (\nu^c \sigma^\mu \bar e^c)  \big |^2 
 \to
 V_{ud }C_{\phi e \nu}  
  (\bar u \bar \sigma_\mu d) (e^c \sigma^\mu \bar \nu^c)   + \hc 
 \end{equation}   
At the end of the day, the tree-level matching between the operators in \cref{eq:EFT_scenario4} and the $\nu$SMEFT Wilson coefficients reads 
 \begin{align}
 \tilde \epsilon_L   = & -   {v^2 \over 2 } C_{\phi e \nu}  , 
 \nnl 
 \tilde \epsilon_R   = & - {v^2 \over 2 V_{ud}} C_{e \nu u d}.  
 \end{align}  
In this scenario, the D-parameter expressed by $\nu$SMEFT Wilson coefficients reads
 \begin{equation}
  \label{eq:EFT_Dscenario4}
 D    \approx  - \kappa_D \,  {v^4 \over 4}   
 \im  \big [ C_{e \nu u d}  C_{\phi e \nu}^* \big ] , 
 \end{equation} 
 which is once again $\cO(\Lambda^{-4})$. 
An important difference with the other scenario with right-handed neutrinos is that the constraints on $\tilde \epsilon_L$ and  $\tilde \epsilon_R$, or on their $\nu$SMEFT counterparts $C_{\phi e \nu}$ and $C_{e \nu u d}$ are much milder. 
In particular, the LEP-2 constraint on W decay  
${\rm Br}(W \to e \nu) = 0.1071(16)$~\cite{ALEPH:2013dgf} translates to 
$|\tilde \epsilon_L| \leq 0.14$ at 95\% CL. 
On the other hand,  $\tilde \epsilon_R$ can be constrained $\ell+$MET searches at the LHC, like the one performed recently by the CMS collaboration~\cite{CMS:2022yjm}.
Although the CMS analysis does not consider the 4-fermion operator in \cref{eq:EFT_scenario4}, we can get an idea about the order of magnitude of the constraint by reinterpreting their limit on sequential $W'$: $m_{W'} \gtrsim 5.4$~TeV in the $e \nu$ decay channel. 
Integrating out the sequential $W'$ leads to the 4-fermion operator 
$(\bar u \bar \sigma^\mu d)(\bar e \bar \sigma_\mu \nu)$ corresponding to $\epsilon_L = {m_W^2\over m_{W'}^2}$, thus $|\epsilon_L| \leq 0.015$.  
Since the operator $(e^c \sigma^\mu \bar \nu^c)(u^c \sigma_\mu \bar d^c)$, unlike the one above, does not interfere with the SM amplitudes, we do not expect it to be more strongly constrained, hence $|\tilde \epsilon_R| \lesssim \cO(0.01)$ is a reasonable estimate. 
All in all $|\tilde \epsilon_{L}| |\tilde \epsilon_{R}| \sim 10^{-3}$ is consistent with the existing bounds. 
However, once again there are stronger constraints on the imaginary part of $\tilde \epsilon_L \tilde \epsilon_R^*$ due to EDMs. 
When both operators in \cref{eq:EFT_scenario4} are present simultaneously, the 4-quark operators in \cref{eq:EFT_C1LR} is generated at one loop. 
While the relevant diagram is quadratically divergent, and thus the contribution to EDM is not calculable within the EFT, one can estimate 
\begin{equation}
 \label{eq:EFT_scenario4C1LR}
C_{1LR} \sim {\Lambda^2 \over 4\pi^2}  C_{e \nu u d}  C_{\phi e \nu}^* .   
\end{equation}
Much as in scenario I, the phase of this operator is correlated with the one responsible for the D parameter, which can be written as 
\begin{equation}
 \label{eq:EFT_scenario4Dest}
D \sim  \kappa_D  10^{-4} 
\bigg ( {v^2  \im  [   C_{1LR} ]   \over 10^{-5}} \bigg )  {v^2 \over \Lambda^2 }. 
\end{equation}  
Therefore $|D|$ can be $\cO(10^{-4})$ only if new physics is at the electroweak scale, and is suppressed when $\Lambda \gg v$. 
A word of caution is that this conclusion is based on the naive dimensional estimate in \cref{eq:EFT_scenario4C1LR}, and should be verified for specific UV completions in which the EDM contribution is actually calculable.

We close this discussion by mentioning one more possibility for realizing this scenario. The WEFT parameter $\tilde \epsilon_L$ can be generated from the dimension-8 operator
\begin{equation}
 \label{eq:EFT_scenario4dim8}
     {\cal L}_{\nu \rm SMEFT} \supset 
     \tilde C_8 (e^c \sigma^\mu \bar \nu^c) (\bar q \tilde H \bar \sigma_\mu H^\dagger q) , 
\end{equation}
instead from the dimension-6 operator in \cref{eq:EFT_scenario4} proportional to $C_{\phi e \nu}$. This leads to 
\begin{equation}
\tilde \epsilon_L = - { v^4 \tilde C_8 \over 4V_{ud} }, 
\end{equation}
and the $D$ parameter reads 
\begin{equation}
 \label{eq:EFT_scenario4bD}
 D    \approx  - \kappa_D \,  {v^6 \over 8 }   
 \im  \big [ C_{e \nu u d}  \tilde C_8^* \big ] . 
\end{equation}
This is $\cO(\Lambda^{-6})$, therefore it is even more strongly suppressed than in other scenarios.  
The reason we consider this option is that $C_{\phi e \nu}$ cannot be generated from models with only leptoquarks and right-handed neutrinos as BSM particles, 
whereas $\tilde C_8$ can. 
We will consider one such leptoquark model in \cref{sec:dr}. 
At the EFT level, the dangerous  $C_{1LR}$ Wilson coefficient is still generated with the quadratically divergent coefficient: 
$C_{1LR} \sim {v^2 \Lambda^2 \over 4\pi^2}  C_{e \nu u d} \tilde C_8^*$.
Consequently, the correlation between the D parameter and the 1-loop-generated $C_{1LR}$ remains the same as in \cref{eq:EFT_scenario4Dest}. 

\subsection{Summary of EFT analysis} 

\begin{table}[]
    \centering
    \begin{tabular}{c|c|c|c|c|}
Scenario & $\nu$WEFT  & $\nu$SMEFT & order $D$ & max $|D|$   \\ \hline
Ia    & $\epsilon_R$ & $H D_\mu H u^c \sigma^\mu \bar d^c$  & 
$\Lambda^{-2}$  & $\cO(10^{-6}) $ \\ 
Ib    & $\epsilon_R$ & $(\bar l H \bar \sigma_\mu H l) (u^c  \sigma^\mu \bar d^c)$ & 
$\Lambda^{-4}$ & $\cO(10^{-4})  {v^2 \over \Lambda^2} $ \\ 
II       & $\epsilon_S$, $\epsilon_T$  & 
$(\bar l \bar \sigma_{\mu \nu} \bar e^c) (\bar q \bar  \sigma^{\mu \nu}  \bar u^c)$, 
$(\bar l \bar e^c)(\bar q \bar u^c)$, 
$(\bar l \bar e^c)(d^c q)$  & 
$\Lambda^{-4}$  & $\cO(10^{-14})$  \\ 
III       & $\tilde \epsilon_S$, $\tilde \epsilon_T$ &
$(\bar l \bar \sigma^{\mu\nu} \bar \nu^c)(\bar q \bar \sigma_{\mu\nu}  \bar d^c)$, 
$(\bar l \bar \nu^c)( \bar q \bar d^c )$, 
$(\bar l \bar \nu^c) (u^c q)$  &
$\Lambda^{-4}$  & $\cO(10^{-6})$  \\ 
IVa      & $\tilde \epsilon_L$, $\tilde \epsilon_R$ & 
$ H^\dagger D_\mu H^\dagger e^c \sigma^\mu \bar \nu^c$, 
$(e^c \sigma^\mu \bar \nu^c) (u^c \sigma_\mu \bar d^c)$ & 
$\Lambda^{-4}$  & $\cO(10^{-4}) {v^2 \over \Lambda^2}$  \\ 
IVb     & $\tilde \epsilon_L$, $\tilde \epsilon_R$ & 
$e^c \sigma^\mu \bar \nu^c \bar q H^\dagger \sigma_\mu H^\dagger q$, 
$(e^c \sigma^\mu \bar \nu^c) (u^c \sigma_\mu \bar d^c)$ & 
$\Lambda^{-6}$ & $\cO(10^{-4}) {v^2 \over \Lambda^2}$  \\ 
    \end{tabular}
    \caption{
Classification of EFT scenarios for generating BSM contributions to the $D$ parameter. 
We list the $\nu$WEFT parameters below the electroweak scale and the $\nu$SMEFT operators above the electroweak scale  that define each scenario. 
We also give the order in the $\nu$SMEFT  EFT expansion parameter $\Lambda$ at which the $D$ parameter appears. 
Finally, we give an estimate of the maximum magnitude of the BSM $D$ parameter in each scenario based on purely EFT and naturalness arguments. }
    \label{tab:scenarios} 
\end{table}

To wrap up our EFT discussion, working within the $\nu$SMEFT extension of the SM, we have classified the scenarios leading to BSM contributions to the $D$ parameter. 
A concise summary is given in \cref{tab:scenarios}, where we list the $\nu$SMEFT operators above the electroweak scale and the $\nu$WEFT parameters below the electroweak scale that define each scenario. 
We also give the maximum magnitude of the BSM $D$ parameter in each scenario, based on purely EFT arguments.

We have identified three interesting scenarios where the D parameter may be at the currently observable level of $\cO(10^{-4})$ without conflicting other experimental data and without fine-tuned cancellations between different EFT Wilson coefficients: 
\begin{enumerate}
    \item {\bf Scenario Ib}, where the D parameter is generated via the 
    $\im \epsilon_R$ term in \cref{eq:EFT_Dquark}, and $\epsilon_R$ descends from the dimension-8 operator  $(\bar l H \bar \sigma_\mu H \tilde l)   (u^c  \sigma^\mu \bar d^c)$ in the $\nu$SMEFT. 
    \item  {\bf Scenario IVa}, where the D parameter is generated via the 
    $\im [\tilde \epsilon_R \tilde \epsilon_L^*] $ term in \cref{eq:EFT_Dquark}, and $\tilde \epsilon_X$ descend from the dimension-6 operators 
    $(e^c \sigma^\mu \bar \nu^c)(u^c \sigma_\mu \bar d^c)$ and 
    $H^\dagger D_\mu H^\dagger (e^c \sigma^\mu \bar \nu^c)$ in the $\nu$SMEFT.
    \item  {\bf Scenario IVb}, similar to the above, except that the dimension-6 operator $H^\dagger D_\mu H^\dagger (e^c \sigma^\mu \bar \nu^c)$ is replaced by the dimension-8 one $(e^c \sigma^\mu \bar \nu^c) (\bar q H^\dagger \sigma_\mu H^\dagger q)$.
\end{enumerate}
In these scenarios, the $D$ parameter of order $10^{-4}$ is generically consistent with other experimental bounds, assuming that the new BSM particles have masses close to the electroweak scale. 
Of course, the corollary is that those models that generate 
$D \sim O(10^{-4} - 10^{-5})$ will eventually face constraints from direct searches at the LHC, which have to be studied for each concrete UV completion separately.  
It should be stressed that the maximum $D$ estimates in these scenarios rely on dimensional estimates of loop contributions to EDMs that are quadratically divergent, and thus not calculable within the EFT.    
They should be verified for specific UV completions in which EDMs are calculable.
We will discuss later on how our estimates compare to one-loop calculations in leptoquark models.

Two more scenarios can lead to an $\cO(10^{-4})$ D parameter at the cost of a percent-level fine tuning:
\begin{enumerate}
\item {\bf Scenario Ia}, where the D parameter is generated via the 
    $\im \epsilon_R$ term in \cref{eq:EFT_Dquark}, and $\epsilon_R$ descends from the dimension-6 operator  $\tilde H^\dagger D_\mu H (u^c \sigma^\mu \bar d^c)$. 
\item  {\bf Scenario III}, where the D parameter is generated via the 
    $\im [\tilde \epsilon_T \tilde \epsilon_S^*] $ term in \cref{eq:EFT_Dquark},
    and $\tilde \epsilon_X$ descend from the scalar dimension-6 operators  
$(\bar l \bar \nu^c)( \bar q \bar d^c )$ and/or  $(\bar l \bar \nu^c) (u^c q )$ together with the tensor one 
$(\bar l \bar \sigma^{\mu\nu} \bar \nu^c)(\bar q \bar \sigma_{\mu\nu}  \bar d^c)$.   
\end{enumerate}  
Finally, in {\bf Scenario II}, where the $D$ parameter is generated via the $\im [\epsilon_T \epsilon_S^*]$ term in \cref{eq:EFT_Dquark}, an enormous fine-tuning would be needed to push the $D$ parameter to observable levels.

In the next section we will discuss which of these scenarios can arise in the BSM models with leptoquarks.

\section{D parameter in CP-violating models with leptoquarks} 
\label{sec:dr}

The general  interactions of leptoquarks with the SM matter are summarized in \cref{sec:lepto}.
In this section we discuss concrete BSM model containing one or two  relatively light leptoquarks contributing to the $D$ parameter. 
We will determine how large can the $D$ parameter be taking into account the existing constraints from high- and low-energy experiments (see e.g.~\cite{deBlas:2013qqa,Crivellin:2021bkd,Allwicher:2022gkm}). 

\subsection{$S_1$-$R_2$} 
\label{sec:s1r2}

We consider a model with two scalar leptoquarks $S_1$ and $R_2$ in the standard nomenclature reviewed in \cref{sec:lepto}. 
Their possible Yukawa  interactions with the SM fermions are summarized in \cref{eq:LQ_L}. 
For the sake of this subsection we set the $S_1$ coupling to right-handed neutrinos to zero, $y_{S d \nu} = 0$, as it is not relevant for the $D$ parameter calculation when $\tilde R_2$ is absent.   
The parameter space is therefore characterized by four complex Yukawa couplings defined in \cref{eq:LQ_Lscalar}: $y_{Sue}$, $y_{Sql}$, $y_{Rqe}$, $y_{Rul}$, 
and two masses $M_{S_1}$ and $M_{R_2}$. 
This model is a realization of the scenario II where the $D$ parameter is generated through the $C_{lequ}^{(1),(3)}$ ($\nu$)SMEFT Wilson coefficients. 
Using the matching in \cref{eq:LQ_matching} the $D$ parameter is expressed by the BSM parameters as
\begin{equation}
D \approx 0.05 \kappa_D  {v^4 \over  M_{S_1^2} M_{R_2}^2 }
 \im \Big [  y_{S q l }  \bar  y_{S u e}   y_{R q e}   \bar y_{R u l}    \Big ]   . 
  \end{equation} 
Clearly, all the four Yukawa couplings have to be non-zero to generate the $D$ parameter, 
and at least one of them should have an imaginary part. 
However, such imaginary parts are prohibitively constrained by EDMs, as discussed around \cref{eq:EFT_scenario2EDM}. 
The imaginary parts of the leptoquark Yukawa couplings are constrained by EDM measurements in the ThO molecule as 
\begin{align}
\bigg |\im [ y_{S u e} \bar y_{S q l } ]  {v^2 \over M_{S_1}^2 }   
+  \im [  y_{R q e}  \bar  y_{R u l}] {v^2  \over  M_{R_2}^2} \bigg |    
\lesssim  & 2 \times 10^{-10}, 
\nnl 
\bigg |\im [ y_{S u e} \bar y_{S q l } ]  {v^2 \over M_{S_1}^2 }   
-  \im [  y_{R q e}  \bar  y_{R u l}] {v^2  \over  M_{R_2}^2} \bigg | 
 \lesssim &  4 \times 10^{-10}. 
\end{align}
Assuming for example that the imaginary part resides in the $S_1$ interactions, we have 
$|D| \lesssim 10^{-11} \kappa_D  {y_{R q e} y_{R u l} v^2 \over  M_{R_2}^2 }$. Furthermore, the masses and Yukawa's of $R_2$ are subject to constraints from the LHC $pp \to e^+ e^-$ process,  ${y_{R q e} y_{R u l} v^2 \over  M_{R_2}^2 } \lesssim 10^{-3}$. 
We conclude that $|D| \lesssim 10^{-14}$ in the $S_1$-$R_2$ model, which is of course  is too small to ever be observed.

Models with the $U_1$-$S_1$, $U_1$-$R_2$, $V_2$-$S_1$, or $V_2$-$R_2$ leptoquark pair can also lead to scenario II, with exactly the same problem due to the EDMs.

\subsection{$S_1$-$\tilde R_2$} 
\label{sec:s1tr2}

We move to a model with two scalar leptoquarks:  $S_1$ and $\tilde R_2$. 
Their quantum numbers are given in \cref{tab:LQ_spin0}, and  their possible Yukawa  interactions with the SM fermions are collected in \cref{eq:LQ_Lscalar}. 
In this model contributions to the D parameter will enter via interactions with right-handed neutrinos, therefore this time we assume $y_{S d \nu}$ is non-zero. 
On the other hand, we set $y_{Sue} = 0$, as this coupling is not relevant for the discussion of the D parameter in this model (and only would make precision constraints more stringent).   
The parameter space is characterized by four complex Yukawa couplings defined in \cref{eq:LQ_Lscalar}: $y_{Sd\nu}$, $y_{Sql}$, $y_{Rq\nu}$, $y_{Rdl}$, 
and two masses $M_{S_1}$ and $M_{\tilde R_2}$. 
This model is a realization of the scenario III where the D parameter is generated through the $C_{l\nu q d}^{(1),(3)}$ $\nu$SMEFT Wilson coefficients. 
Using the matching in \cref{eq:LQ_matching} the D parameter is expressed by the BSM parameters as 
\begin{equation}
D \approx 0.05   \kappa_D { v^4 \over M_{S_1}^2 M_{\tilde R_2}^2 }
 \im \big [ y_{S d \nu} \bar y_{S q l }  \bar y_{R q \nu}  y_{R d l}    \big ]  . 
  \end{equation}  
The magnitude of Yukawa couplings entering this formula are constrained by precision measurements of CP conserving quantities, notably by pion decay. 
Indeed, in this model the dangerous Wilson coefficient $\tilde \epsilon_P$ is generated in the WEFT below the electroweak scale. 
At tree level and ignoring the (nearly identical for $\tilde{\epsilon}_S$ and $\tilde{\epsilon}_P$) running effects in \cref{eq:EFT_Lrweft} one finds 
\begin{equation}
\tilde \epsilon_P  =   - {v^2 \over 4 } \bigg [
{y_{S d \nu} \bar y_{S q l }   \over  M_{S_1}^2}  
   +   {y_{R q \nu} \bar y_{R d l}       \over  M_{\tilde R_2}^2}
\bigg ],  
  \end{equation}  
  Thus,
  \begin{equation}
|D|  \lesssim  0.2 |\kappa_D| |\tilde \epsilon_P| 
 {\rm Min} \bigg [ 
|y_{S d \nu} y_{S q l }|  { v^2 \over  M_{S_1}^2}  , 
|y_{R q \nu}  y_{R d l}| { v^2  \over M_{\tilde R_2}^2 }   \bigg ]  . 
  \end{equation}  
The pion decay bound $|\tilde \epsilon_P| < 1.0 \times 10^{-5}$ implies  $|D| < 10^{-5}$ for order one Yukawa couplings and leptoquark masses at the electroweak scale. 
In this model there are no free parameters that could be used to fine-tune away the pion decay constraints on the $D$ parameter: if $\tilde \epsilon_P$ is fine-tuned to vanish, so does the $D$ parameter. 
Taking into account the LHC constraints, the leptoquark contributions to the $D$ parameter are further suppressed. 
The constraints from the $pp \to e^+ e^-$ Drell-Yan process imply 
${|y_{S q l }|^2  v^2 \over  M_{S_1}^2}, { v^2 |y_{R d l}|^2 \over M_{\tilde R_2}^2 } \lesssim 10^{-3}$.   
While the constraints from the $pp \to e^+ \nu$ are weaker,  ${|y_{S d \nu}|^2  v^2 \over  M_{S_1}^2}, { v^2 |y_{R q \nu}|^2 \over M_{\tilde R_2}^2 } \lesssim 10^{-1}$, 
overall $|D| \lesssim 10^{-7}$. 
This is too small to be observable in any foreseeable future.

Models with the $U_1$-$S_1$, $U_1$-$\tilde R_2$, $\tilde V_2$-$S_1$, or $\tilde V_2$-$\tilde R_2$ leptoquark pair can also lead to scenario III.  
They have very similar properties as the $S_1$-$\tilde R_2$ model, and only differ by order one factors regarding the constraints.  
In models with at least three leptoquarks, for example   $U_1$-$S_1$-$\tilde R_2$, there exists a possibility to fine tune away the contributions to $\tilde \epsilon_P$ and somewhat alleviate the problem of the pion decay constraints.

\subsection{$R_2$-$\tilde R_2$ } 
\label{sec:ng} 

\begin{figure}[tb]
    \centering
    \includegraphics[width=0.45\textwidth]{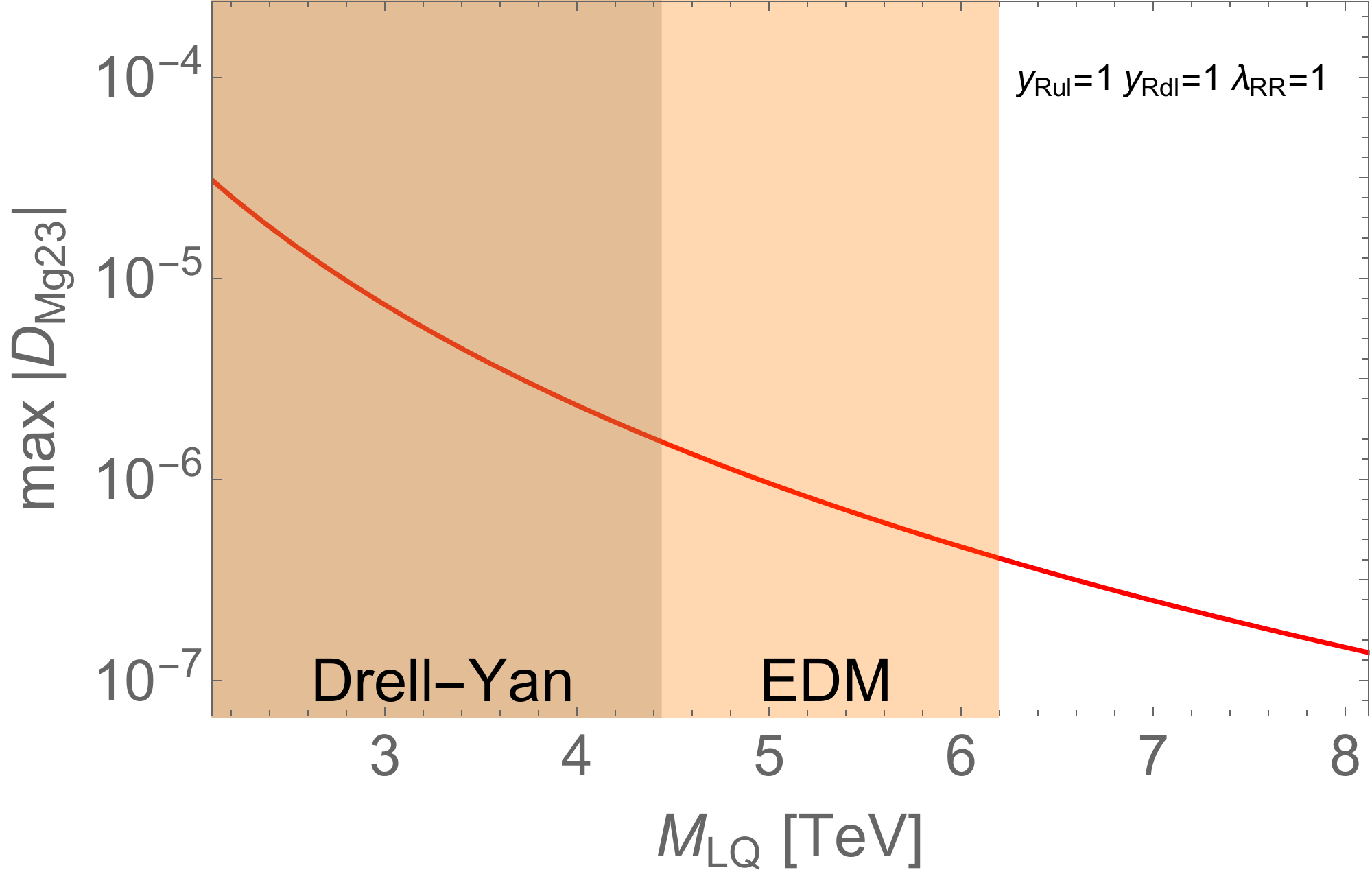}
    \quad
\includegraphics[width=0.45\textwidth]{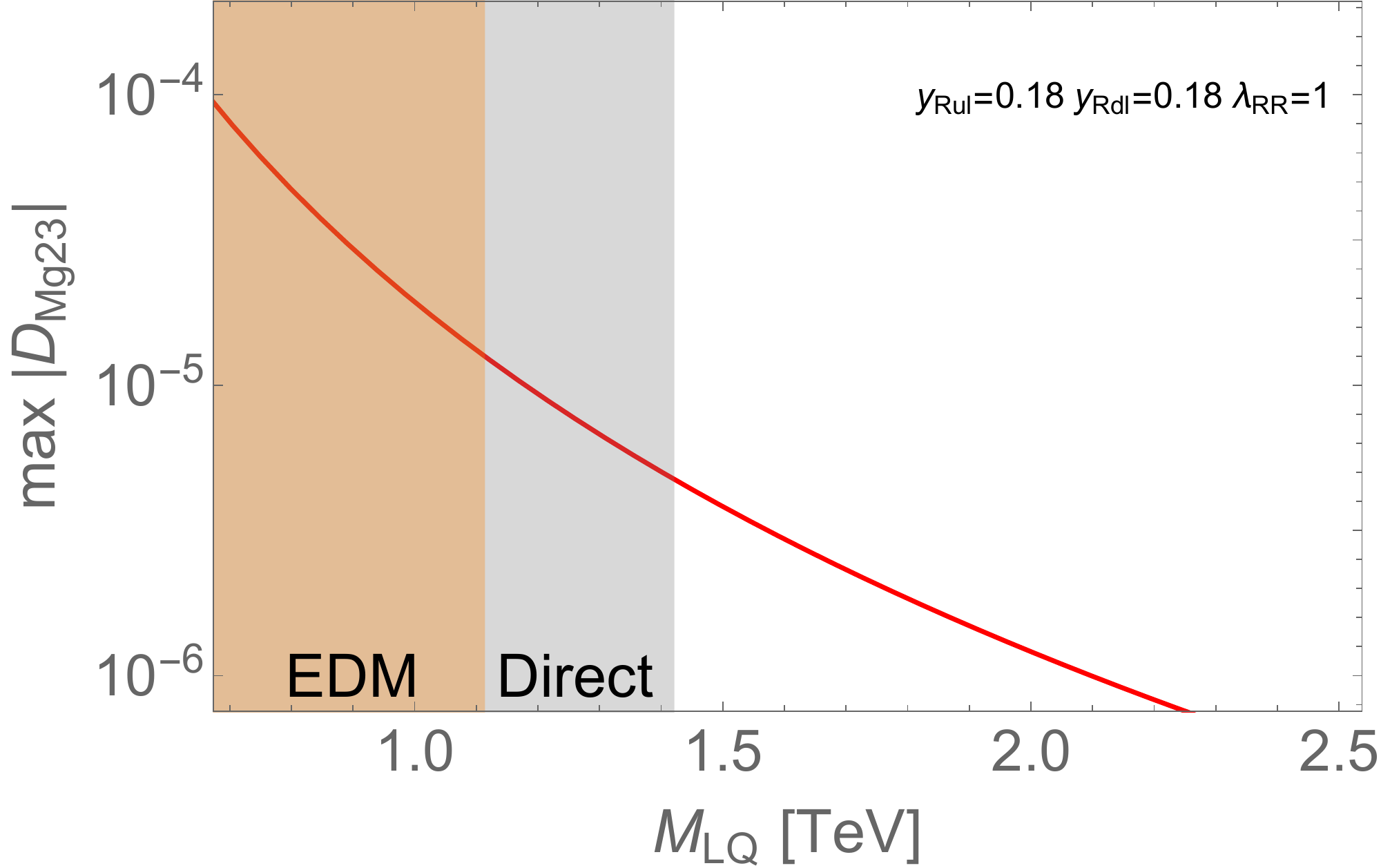}
    \caption{Maximum value of the $D$ parameter in ${}^{23}$Mg (red line) possible in the model with two leptoquarks $R_2$ and $\tilde R_2$ for two particular choices of the leptoquark Yukawa couplings and for the scalar mixing parameter $\lambda_{RR}=1$.  
    Left: for $y_{Rul} = y_{Rdl} = 1$ the colored parameter space is excluded by $pp \to e^+ e^-$ (Drell-Yan)~\cite{Allwicher:2022mcg} and by the neutron EDM. The latter dominate and force the leptoquark mass scale to be at least $\sim 6$~TeV, which translates into a strong suppression of the $D$ parameter to the level below $4 \times 10^{-7}$. Right: for $y_{Rul} = y_{Rdl} = 0.18$ we show the constraints  from  pair production at the LHC (direct) which dominate over the Drell-Yan (not shown) and EDM constraints. This is a more favorable situation from the point of view of the $D$ parameter, allowing for $|D| \approx 5 \times 10^{-6}$. }
    \label{fig:R2tR2}
\end{figure}

We turn to the model with one $R_2$ and $\tilde R_2$, which was already discussed in Refs.~\cite{Herczeg:2001vk,Ng:2011ui}, and leads to scenario Ib in the nomenclature of \cref{tab:scenarios}.  
In order to generate the $D$ parameter we need the Yukawa couplings $y_{Rul}$ and $ y_{Rdl}$ in \cref{eq:LQ_Lscalar} (and also the quartic mixing in \cref{eq:LQ_RRmixing}) to be non-zero. 
On the other hand, in this subsection we set $y_{Rqe} = y_{Rq\nu} = 0$ for simplicity. 
The $D$ parameter is given by 
\begin{equation}
 D    \approx  - {\kappa_D v^4 \over 8  M_{R_2}^2 M_{\tilde R_2}^2}   
 \im  \big [ \lambda_{RR} y_{Rul} \bar y_{R d l}  \big ].   
\end{equation}
 We will assume that the phase of $\lambda_{RR} y_{Rul} \bar y_{R d l}$ is maximal. 
In this case, the possibility of generating a large $D$ parameter becomes severely constrained when combining existing experimental constraints. 
We begin with EDMs. 
In this model, the 4-quark operator in \cref{eq:EFT_C1LR} is generated by leptoquark loops~\cite{Ng:2011ui} with the same phase as the one entering the $D$ parameter. 
Unlike in the EFT calculation discussed in \cref{sec:EFT}, the loop is finite and therefore calculable without ambiguities, since the model is renormalizable. 
For leptoquark masses sufficiently larger than the electroweak scale one finds~\cite{Ng:2011ui}
\begin{equation}
\im C_{1LR} \approx
\im  \big [ \lambda_{RR} y_{Rul} \bar y_{R d l}  \big ] 
{\log(M_{R_2}^2/M_{\tilde R_2}^2) \over 16 \pi^2(M_{R_2}^2 - M_{\tilde R_2}^2)} . 
\end{equation}
This is always a good approximation for leptoquarks satisfying the direct search constraints (see below). Note that 
$\lim_{x \to y} \log(x/y)/(x-y) = 1/y$. 
We can rewrite the $D$ parameter as 
\begin{equation}
 D    \approx  - 2 \pi^2 \kappa_D    \im C_{1LR} 
{v^4 (M_{R_2}^2 - M_{\tilde R_2}^2) \over  M_{R_2}^2 M_{\tilde R_2}^2 \log(M_{R_2}^2/M_{\tilde R_2}^2) }. 
\end{equation} 
The most favorable situation for the $D$ parameter corresponds to the limit 
$M_{\tilde R_2}^2 = M_{R_2}^2 \equiv M_{\rm LQ}^2$. 
We thus have at the inequality 
\begin{equation}
\label{eq:R2tR2-Dinequality}
|D|    \lesssim  2 \times 10^{-4} |\kappa_D|    {v^2 |\im C_{1LR}|  \over 10^{-5} }  {v^2 \over M_{LQ}^2} . 
\end{equation} 
The LHC constraints from leptoquark pair production leading to the $qq l l$ final state~\cite{CMS:2018ncu} imply $M_{R_2,\tilde R_2} \gtrsim 1.4$~TeV~\cite{Crivellin:2021egp}, independently to a large extent of the value of the Yukawa couplings. 
Thus $v^2/M_{LQ}^2 \lesssim 3 \times 10^{-2}$ and we arrive at the bound 
\begin{equation}
\label{eq:R2tR2-maxD}
|D|    \lesssim 6 \times 10^{-6} |\kappa_D| . 
\end{equation}
in the entire phenomenologically allowed parameter space of the $R_2$-$\tilde R_2$ model.  
For a given Yukawa coupling, further constraints can be derived using the leptoquark contributions to the Drell-Yan process $pp \to e^+ e^-$. 
Here we will work in the limit where $M_{R_2,\tilde R_2} \gg 1$~TeV such that the Drell-Yan process can be accurately described within SMEFT. 
As shown in Eqs.~(\ref{eq:LQ_LeffCPconserving}) and (\ref{eq:LQwilsons}), 
integrating out the leptoquarks generates the effective 4-fermion operators 
\begin{align}
{\cal L}_{\nu \rm SMEFT } \supset  &
 - {  |y_{R u l}|^2 \over 2 M_{R_2}^2}    (\bar l \bar \sigma^\mu l)  (u^c  \sigma^\mu \bar u^c ) 
-    { |y_{R d l}|^2   \over 2 M_{\tilde R_2}^2 }    (\bar l \bar \sigma^\mu l)  (d^c  \sigma^\mu \bar d^c ), 
\end{align}
which contribute to $pp \to e^+ e^-$. 
We  use the bounds from the analysis of Ref.~\cite{Allwicher:2022mcg} based on the CMS and ATLAS $e^+ e^-$ pair production results~\cite{CMS:2021ctt,ATLAS:2019lsy}. 
The Drell-Yan, pair production, and  EDM constraints together are shown in \cref{fig:R2tR2} for two particular choices of the Yukawa couplings and the scalar mixing $\lambda_{RR}$. 
For large Yukawas, the EDM bounds push the leptoquark mass scale into the multi-TeV regime, leading to a stronger suppression of the $D$ parameter than the maximum value on \cref{eq:R2tR2-maxD}. 
In this regime, the Drell-Yan bounds are stronger than the pair production ones, but always weaker than the EDM ones. 
Very small Yukawas also suppress the $D$ parameter, and in this case the limit is set by the direct bounds on leptoquark masses from pair production (which does not depend on $y_{Rul}$ and $y_{Rdl}$). 
The $D$ parameter is maximized when the EDM and direct bounds coincide, 
which happens for $|y_{Rul}| \approx  |y_{R d l}| \approx 0.2$. 
For these sweet-spot values of the Yukawas, the Drell-Yan bounds happen to be somewhat weaker than the pair production ones, due to a small $\sim 2\sigma$ excess in the former data. 
All in all, we find that maximum $|D| \approx 8 \times 10^{-6}$ is possible for ${}^{23}$Mg. 
This is still below the sensitivity of the next generation experiments, but it could be a realistic goalpost in the future.

\subsection{$R_2$-$\tilde R_2$-$S_1$ } 
\label{sec:ng2} 

\begin{figure}[tb]
    \centering
    \includegraphics[width=0.45\textwidth]{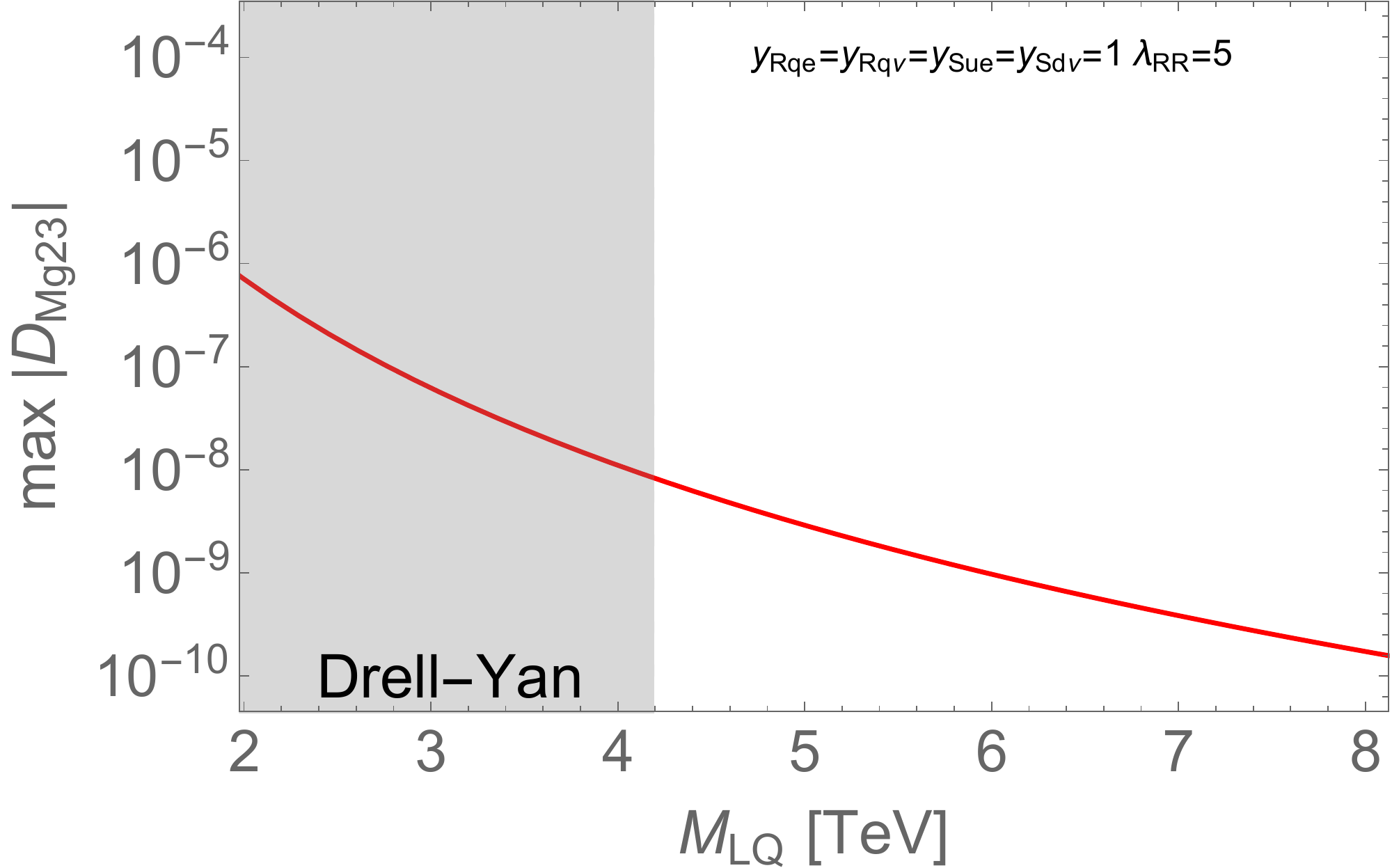}
    \quad
\includegraphics[width=0.45\textwidth]{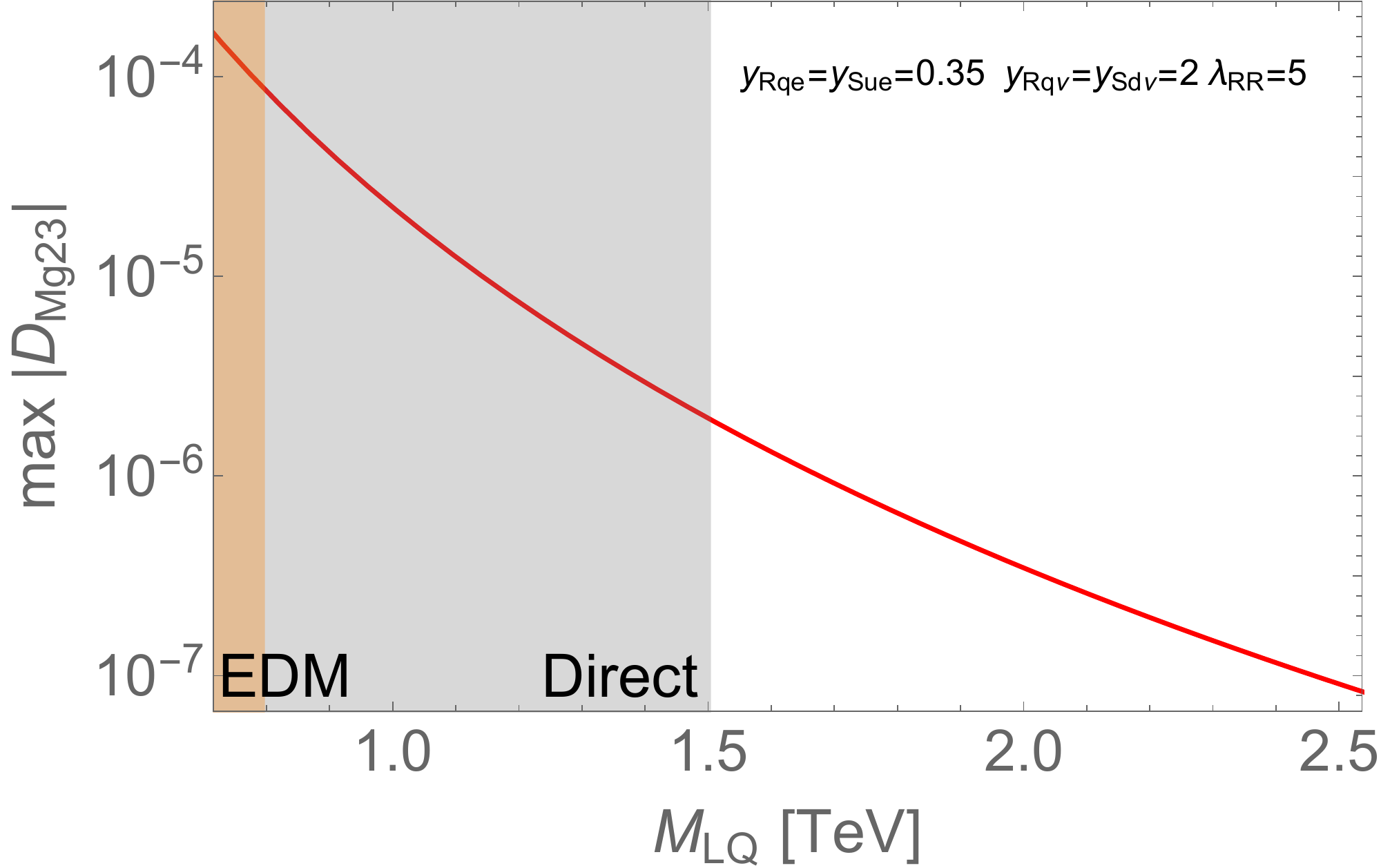}
    \caption{Maximum value of the $D$ parameter in ${}^{23}$Mg (red line) possible in the model with two leptoquarks $R_2$ and $\tilde R_2$ for two particular choices of the leptoquark Yukawa couplings and for the scalar mixing parameter $\lambda_{RR}=5$.  
  Left: for $|y_{S d \nu}| = |y_{S u e}|=  |y_{R q e}| = |y_{R q \nu}| = 1$ the gray-shaded area (Drell-Yan) is excluded by $pp \to e^+ e^-$~\cite{Allwicher:2022mcg}. 
  Due to the strong Drell-Yan bound on $M$ the $D$ parameter is suppressed below $\cO(10^{-8})$ in the allowed parameter space.  
 For these Yukawa couplings  the EDM constraints are much weaker, $M_{\rm LQ} \lesssim 1$~TeV. 
 Right: for $ |y_{R q e}|  = |y_{S u e}|=  0.35$ and $|y_{S d \nu}| = |y_{R q \nu}| = 2$,  one can arrive at  $|D| \approx 2 \times 10^{-6}$ in the allowed parameter space. 
The gray-shaded area (Direct) is excluded by the LHC searches for leptoquark pair production. For these Yukawa couplings  $pp \to e^+ e^-$ gives almost the same exclusion limit, although for such moderate values of $M$ the EFT analysis we perform is not a very good approximation of the Drell-Yan process in the full leptoquark model and it overestimates the bounds somewhat.  
The orange-shaded are shows the EDM constraints, which are again weaker. 
    \label{fig:R2tR2S1}
    }
\end{figure}

The $R_2$-$\tilde R_2$-$S_1$ model considered in this subsection was discussed in Ref.~\cite{Ng:2011ui}, and it realizes scenario IVb  in the nomenclature of \cref{tab:scenarios}.  
While the dimension-6 $\nu$SMEFT operator $H^\dagger D_\mu H^\dagger e^c \sigma_\mu \bar \nu^c$ cannot be generated at tree level in models where the leptoquarks are the only exotic particles (for that one would have to extend the gauge symmetry of the SM and introduce the associated gauge bosons),  the dimension-8 $\nu$SMEFT operator $e^c \sigma_\mu \bar \nu^c \bar q \tilde H \sigma_\mu H^\dagger q$ can be generated by the $R_2$-$\tilde R_2$ pair mixing after electroweak symmetry breaking. 
In this subsection we set the Yukawa couplings $y_{Rul}=y_{Rdl}=0$ in \cref{eq:LQ_Lscalar}, as they will not play any role in generating $D$. 
On the other hand, this time we assume that $y_{Rqe}$ and $y_{Rq\nu}$ are non-zero. 
Integrating out the $R_2$-$\tilde R_2$ pair in the presence of the scalar mixing in \cref{eq:LQ_RRmixing} generates the $e^c \sigma_\mu \bar \nu^c \bar q \tilde H \sigma_\mu H^\dagger q$ operator with the Wilson coefficient 
\begin{equation}
\tilde C_8 = - {\bar \lambda_{RR} \bar y_{R q e} y_{R q \nu} \over 2 M_{R_2}^2 M_{\tilde R_2}^2 } . 
\end{equation}
In scenario IV, one also has to generate the Wilson coefficient  $C_{e \nu u d}$ of the dimension-6 operator 
$e^c \sigma_\mu \bar \nu^c u^c \sigma_\mu \bar d^c $.
To this end we introduce in addition the $S_1$ leptoquark, and assume its Yukawa couplings $y_{Sue}$ and $y_{Sd\nu}$ in \cref{eq:LQ_Lscalar} are non-zero.
In this subsection we set $y_{Sql} =0$ for simplicity.
From Eq.~(\ref{eq:LQ_matching}) one then has
\begin{equation}
 C_{e \nu u d} =     { y_{S d \nu} \bar y_{S u e}   \over 2 M_{S_1}^2} . 
\end{equation}
Plugging the above Wilson coefficients into \cref{eq:EFT_scenario4bD} one finds the $D$ parameter 
\begin{equation}
 D    \approx   \kappa_D \,  {v^6 \over 32 M_{S_1}^2  M_{R_2}^2 M_{\tilde R_2}^2}    \im  \big [  \lambda_{RR} y_{S d \nu} \bar y_{S u e}  y_{R q e} \bar y_{R q \nu} \big ] . 
\end{equation}
In this scenario the $D$ parameter is suppressed by six powers of leptoquark masses, which then have to be very close to the electroweak scale to obtain a sizable magnitude of $D$. 
Therefore one expects that a large $D$ will be in tension with direct LHC searches.
The LHC constraints from leptoquark pair production imply $M_{R_2,\tilde R_2,S_1} \gtrsim 1.4$~TeV~\cite{Crivellin:2021egp}, independently to a large extent of the value of the Yukawa couplings.
This alone brings a suppression factor of $\cO(10^{-5})$, which would have to be balanced by large Yukawas and/or scalar mixing. 
Further constraints come from EDMs generated at one loop. 
In the limit where $M_{R_2} = M_{\tilde R_2}=M_{S_1} \equiv M_{\rm LQ}$ one generates the 4-quark operator in \cref{eq:EFT_C1LR} with the Wilson coefficient~\cite{Ng:2011ui} 
\begin{equation}
|\im C_{1LR}|  = {v^2 \over 256 \pi^2  M_{\rm LQ}^4 }  
\big | \im  \big [  \lambda_{RR} y_{S d \nu} \bar y_{S u e}  y_{R q e} \bar y_{R q \nu} \big ] \big |  + \cO(M_{\rm LQ}^{-6} ) . 
\end{equation}
Therefore the $D$ parameter can be recast as 
\begin{equation}
\label{eq:R2tR2S1-Dinequality}
 |D|    \lesssim  8 \times 10^{-4}  |\kappa_D|  
 {v^2 |\im C_{1LR}| \over 10^{-5}}  {v^2 \over M_{\rm LQ}^2} . 
\end{equation}
The above is an inequality because hierarchies between the leptoquark masses lead to a suppression of $D$ with respect to $\im C_{1LR}$. 
Given the EDM constraints, the maximum value of the $D$ parameter is $4$ times larger than in the model of the previous subsection, cf.~\cref{eq:R2tR2-Dinequality}. 
Taking into account the direct LHC constraints on leptoquarks, at face value one can reach $|D| \approx 2 \times 10^{-5}$.
However, as we discuss below, in the realistic parameter space  $|D|$ is always smaller. One reason is that the present model also faces constraints from $pp \to e^+ e^-$ Drell-Yan production. 
For $M_{\rm LQ} \gg v$ this is described by the effective operators
\begin{align}
\label{eq:S1R2tR2_LeffDY}
{\cal L}_{\nu \rm SMEFT } \supset  &
{| y_{S u e}  |^2  \over 2  M_{S_1}^2} (u^c  \sigma^\mu \bar u^c )  (e^c \sigma^\mu \bar e^c ) 
- {|y_{R q e}|^2  \over 2 M_{R_2}^2}   (\bar q \bar \sigma^\mu q) (e^c \sigma^\mu \bar e^c )  . 
\end{align}
In \cref{fig:R2tR2S1} we contrast the maximum value of $D$ parameter in this model with the direct, Drell-Yan, and EDM constraints.  
For order one  $y_{S u e}$ and $y_{R q e}$, the LHC process $pp \to e^+ e^-$ provides the strongest constraint, pushing the leptoquark mass scale above $4$ TeV, which leads to a prohibitive suppression of $D$. 
A larger $D$ can be achieved in the regime where $y_{S u e}$ and $y_{R q e}$ are somewhat suppressed  while $y_{S d \nu}$ and  $y_{R q \nu}$ are enhanced, in which case the direct LHC constraints from leptoquark pair production dominate. 
Still, one can achieve $|D| \sim 3 \times 10^{-6}$ at best. 
In order to get to the values suggested by \cref{eq:R2tR2S1-Dinequality} one would have to further increase  $y_{S d \nu} y_{R q \nu}$. 
This would not only be at odds with perturbativity, but would also be subject to bounds  from $pp \to e^\pm nu$ production at the LHC, 
$|y_{S d \nu}|^2/M_{\rm LQ}^2, |y_{R q \nu}|^2/M_{\rm LQ}^2 \lesssim 10^{-1}$. 
We conclude that the $R_2$-$\tilde R_2$-$S_1$ model cannot lead to a larger $D$ parameter than  the $R_2$-$\tilde R_2$ one, in spite of more favorable EDM constraints.
The problem is rooted in the fact that $D \sim \cO(M_{\rm LQ}^{-6})$ in the former, and is thus strongly suppressed given the LHC constraints. 
It is reasonable to conjecture that any BSM model with the new physics scale $\Lambda$ where $D \sim \cO(\Lambda^{-6})$ will suffer from similar bounds.\footnote{%
One could also consider a hybrid $S_1$-$W'$ scenario where, instead of $\tilde C_8$, one generates $e^c \sigma_\mu \bar \nu^c \bar q \tilde H \sigma_\mu H^\dagger q$ by integrating out the right-handed $W$ boson. This would realize scenario IVa in the nomenclature of \cref{tab:scenarios}, in which case  $D \sim \cO(\Lambda^{-4})$. 
However, direct LHC constraints on new gauge bosons are typically a factor of few stronger than those on leptoquarks, and it is unlikely that this avenue could lead to a significant enhancement of the $D$ parameter.   
}  

\section{CP-conserving new physics via the D parameter? } 
\label{sec:cpcon}

We have seen in the previous sections that imaginary parts of the Wilson coefficients $C_X^\pm$, of the magnitude that could lead to a potentially observable $D$ parameter,  are strongly constrained by EDMs.
Therefore it is interesting to note that the one-loop electromagnetic corrections to the beta decay amplitude also contribute imaginary parts to the amplitude when some particle in the loop goes on shell. 
These imaginary parts contribute to the $D$ parameter and other correlation coefficients in beta decay~\cite{Jackson:1957auh}.
But they do not contribute to EDMs (since electromagnetic interactions are CP-conserving) and are thus less constrained. 
In this section we argue that the scenario where precision measurements of the $D$ parameter uncover {\em CP-conserving} new physics is perhaps more realistic than observing CP violation in this kind of experiments.

At one loop, the chief electromagnetic effect on the beta decay amplitude arises due to photon exchange between the daughter nucleus and the beta particle.
When these last two particles in the loop are on shell the amplitude develops an imaginary part which, via unitarity, is fully determined by tree level amplitudes. 
The leading term in the non-relativistic expansion of that contribution is referred to as the Coulomb correction and is proportional to 
$\Delta_Z \equiv { m_e \over p_e} Z_{\mathcal{N}'} \alpha$,
where $Z_{\mathcal{N}'}$ is the charge of the daughter nucleus, $\alpha$ is the fine-structure constant, $m_e$ is the electron mass, and $p_e$ is the momentum of the outgoing $\beta$ particle.
In fact, at the leading order in $1/m_N$, the Coulomb corrections to any correlation coefficient can be effectively described by the following transformation of the Wilson coefficients in the leading order expression:  
\begin{align}
\label{eq:LOOP_zRotation}
C_V^\pm  \to  & C_V^\pm   \pm  {i  \Delta_Z  \over 2}   C_S^\pm ,
\qquad 
C_S^\pm  \to   C_S^\pm   \pm  {i \Delta_Z  \over 2}    C_V^\pm , 
 \nnl
  C_A^\pm \to  & C_A^\pm \pm  {i \Delta_Z  \over 2}   C_T^\pm , 
\qquad  
 C_T^\pm \to  C_T^\pm   \pm  {i \Delta_Z  \over 2}   C_A^\pm , 
\end{align}
where the $\pm$ sign in front of  $\Delta_Z$ refers to the $\beta^\mp$ transitions.
See \cref{sec:dloops} for a derivation of \cref{eq:LOOP_zRotation}.
Inserting the shift \cref{eq:LOOP_zRotation} into \cref{eq:EFT_Dnucleon}, the expression for the $D$ parameter is generalized as $D = D_{\rm LO} + D_{\rm Coulomb}$, where $D_{\rm LO}$ is given by the expression in \cref{eq:EFT_Dnucleon}, and the  Coulomb correction is~\cite{Jackson:1957auh} 
\begin{align}
\label{eq:LOOP_Dnucleon}
D_{\rm Coulomb}  = &   \pm  2r \sqrt{J \over J+1}  {\Delta_Z \, \re \big [    C_V^+ \bar C_T^+ - C_S^+ \bar C_A^+  
+ C_V^- \bar C_T^- - C_S^- \bar C_A^- \big ]
\over 
 |C_V^+|^2   + |C_S^+|^2  +  |C_V^-|^2   + |C_S^-|^2  
+  r^2 \big [  |C_A^+|^2  + |C_T^+|^2  +  |C_A^-|^2  + |C_T^-|^2 \big ] }    , 
\end{align}
The important point is that $D_{\rm Coulomb}$ can be non-zero even for real Wilson coefficients, 
that is in the absence of CP violation in the nucleon-level EFT.  
Note that $D_{\rm Coulomb}$ is very small in the SM\footnote{%
In the SM, larger contributions arise due to interference between the Coulomb corrections and Wilson coefficients of interactions subleading in recoil (such as e.g. the weak magnetism).
These are of order $\sim 10^{-4}$ and are known with a certain precision~\cite{Callan:1967zz,Chen:1969hkp,Holstein:1972bbl,Ando:2009jk}.
} where $C_{S}^{-}=C_{T}^{\pm}=0$ and
$C_S^+$ is suppressed by both $m_e/m_p \sim 10^{-3}$ and isospin breaking, 
leading to the estimate $v^2 C_{S, \rm SM}^+ \lesssim 10^{-5}$.   
On   the other hand, in a general CP-conserving BSM set-up $C_{S,T}^{\pm}$ can be non-zero.
The current model independent constraints imply $v^2 \re C_{S,T}^+ \lesssim 10^{-3}$, 
while  $v^2 \re C_{X}^- \lesssim 10^{-1}$~\cite{Falkowski:2020pma}. 
The additional suppression factor of $Z_{\mathcal{N}'} \alpha$   
is not a very small number, 
especially for beta decays with $Z_{\mathcal{N}'}\gtrsim 10$. 
All in all, 
with completely real Wilson coefficients,
it is possible to arrange for $\cO(10^{-5})$ BSM contributions to the $D$ parameter in heavier nuclei,  
or even $\cO(10^{-4})$ when right-handed neutrinos are present in the low-energy EFT.
Taking into account that a sensitivity of $\cO(10^{-4})$ is expected in the near future \cite{Delahaye:2018kwf}, 
CP-conserving new physics is a promising target for experiments measuring $D$.

\begin{figure}
    \centering
    \includegraphics[width=0.5\textwidth]{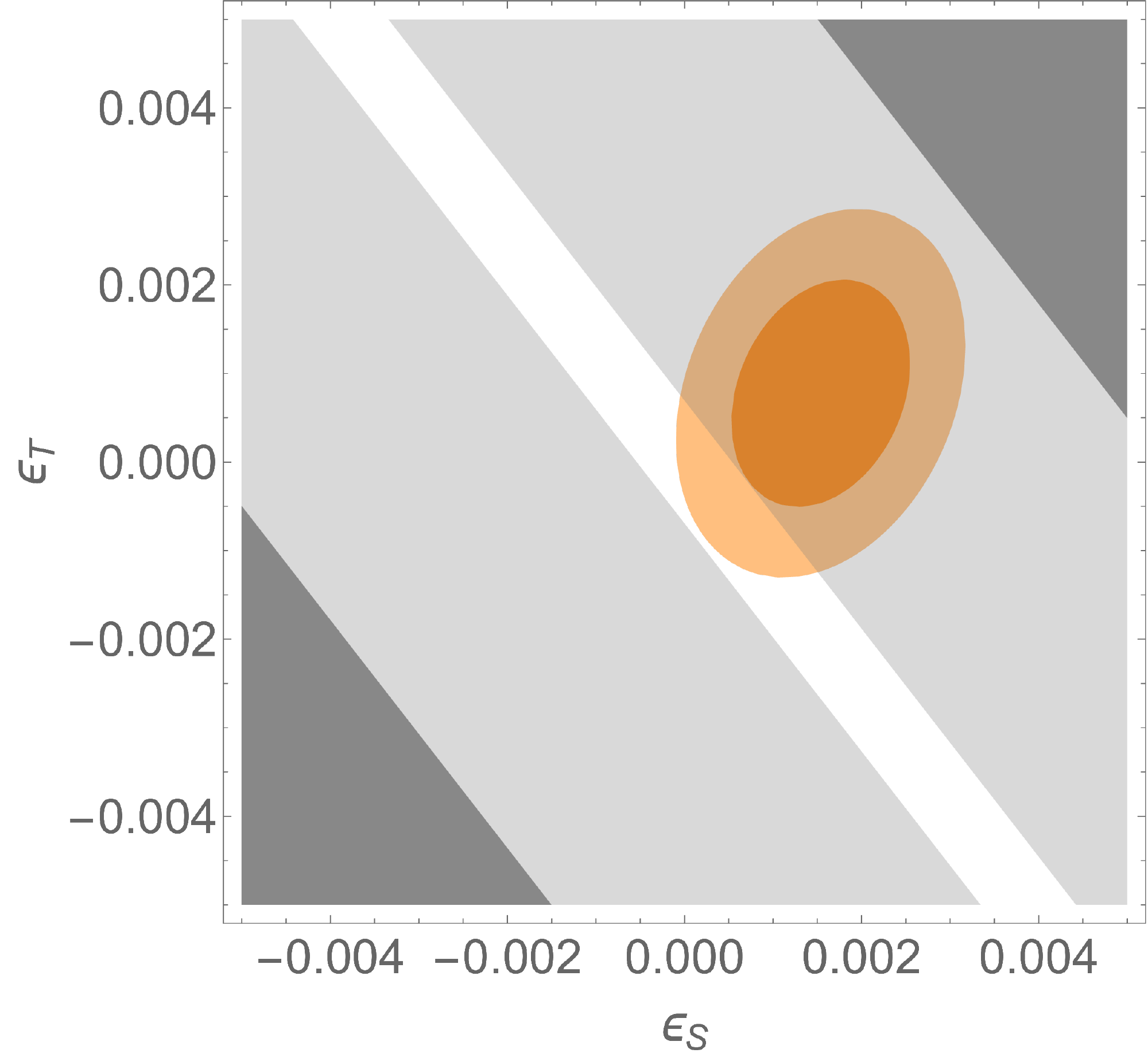}
    \caption{
Constraints on the Wilson coefficients $\epsilon_S$ and $\epsilon_T$ in the quark-level Lagrangian \cref{eq:EFT_Lrweft} at the scale $\mu\simeq 2$~GeV, 
assuming $\tilde \epsilon_X = 0$ and marginalized over the remaining Wilson coefficients. 
We show the region preferred by the combination of low-energy precision measurements performed in Ref.~\cite{Cirigliano:2021yto} at 68\% CL (darker orange), and 95\% CL (lighter orange).  
We also show the region excluded by a hypothetical measurement of the $D$ parameter in the ${}^{23}$Mg beta decay with the uncertainty of $10^{-4}$ (darker shade of gray) and $10^{-5}$ (lighter shade of gray).}
    \label{fig:EFT}
\end{figure}

Translating \cref{eq:LOOP_Dnucleon} to the quark-level EFT parameters one finds
\begin{equation}
D_{\rm Coulomb}\approx \pm  Z_{\mathcal{N}'} \alpha \frac{\kappa_{D}}{2}\frac{m_e}{p_e}\left[ 
\frac{g_T}{g_A}\epsilon_T
+ {g_S \over g_V} \epsilon_S
+\frac{g_T}{g_A}(\tilde{\epsilon}_R+\tilde{\epsilon}_L)\tilde{\epsilon}_T 
+{g_S \over g_V}  (\tilde{\epsilon}_R-\tilde{\epsilon}_L) \tilde{\epsilon}_S
 \right] \, , 
\end{equation}
where $\kappa_D$ is defined in \cref{eq:EFT_Dquark}, and we neglected $\cO(\epsilon_X^2)$ (but not $\cO(\tilde \epsilon_X^2)$) terms.
As a reference, let us take ${}^{23}$Mg, with $Z_{\mathcal{N}'}=11$, $\kappa_{D}\approx -1.3$ and the averaged $\big \langle \frac{m_e}{p_e} \big \rangle \approx 0.35$. 
One finds
\begin{equation}
D_{\rm Coulomb}^{23\mathrm{Mg}}\approx 0.019\left[\epsilon_S+(\tilde{\epsilon}_R-\tilde{\epsilon}_L)\tilde{\epsilon}_S\right]+0.014\left[\epsilon_T+(\tilde{\epsilon}_R+\tilde{\epsilon}_L)\tilde{\epsilon}_T\right] \, .
\end{equation}
Assuming the absence of CP violation, the future sensitivity of $\Delta D^{23\mathrm{Mg}}\sim 10^{-5}$ would translate into a sensitivity on $\epsilon_{S}$ and $\epsilon_{T}$ of about $\sim 10^{-3}$.\footnote{%
The bound on $\epsilon_S$ is perhaps more interesting from the perspective of the UV completion of the EFT. 
Once again, this is a consequence of the mixing between $\epsilon_T$  and $\epsilon_P$ under renormalization group running. 
Namely, $\epsilon_T \sim 10^{-3}$ at the scale $M_Z$ induces $\epsilon_P \sim 10^{-5}$ at $\mu \simeq 2$~GeV~\cite{Gonzalez-Alonso:2017iyc}, 
at odds with the bound  $|\epsilon_P(2~{\rm GeV})| \lesssim 5 \times 10^{-7}$ from $\pi^- \rightarrow e^{-}\bar{\nu}_{e}$. 
Barring  artificial fine tuning between two unrelated effects -  the QED running between two renormalization scales and the exact linear combinations of Wilson coefficients generated from new interactions at higher energies - the pion decay bound on $\epsilon_T(M_Z)$ is stronger than what one can realistically obtain from the  $D$ parameter.}
This is the same order of magnitude as the sensitivity offered by the currently most precise CP-even probes in $\beta$ decays: superallowed $0^+ \to 0^+$ and neutron decays~\cite{Falkowski:2020pma}.
We illustrate this point in \cref{fig:EFT}, where hypothetical constraints on $\epsilon_S$ and $\epsilon_T$ from future measurement of $D^{23\mathrm{Mg}}$ are compared to the existing ones from the combination of low-energy precision measurements performed in Ref.~\cite{Cirigliano:2021yto}.  
The conclusion is that a  measurement of $D$ with an uncertainty better than $10^{-4}$ would already affect the current constraints.
One important point is that the sensitivity of $D$ to the real part of $\epsilon_S$ and $\epsilon_T$ is {\em linear}, and that scales as $1/M^2$ in the new physics scale $M$. This is in contrast to most  CP-violating scenarios considered earlier, where the sensitivity scaled as $1/M^4$, which lead to strong tension with direct LHC bounds. 
If right-handed neutrinos are present in the low-energy EFT, a potential sensitivity of $\Delta D^{23\mathrm{Mg}}\sim 10^{-5}$ would constrain the products of two $\tilde{\epsilon}_X$ at the $10^{-3}$ level. 
This sensitivity is highly competitive with the most precise CP even bounds from $\beta$ decays, which give 
$|\tilde{\epsilon}_X|\lesssim 0.1$~\cite{Falkowski:2020pma}. 

Moving upwards in our EFT ladder, if we assume the validity of the SMEFT at the TeV energies, the potential bound on $\epsilon_S$ and $\epsilon_T$ translates into a bound on the $C_{lequ}^{(1)}$, $C_{lequ}^{(3)}$, and $C_{ledq}$  Wilson coefficients in \cref{eq:EFT_scenario2}. 
Assuming that new physics generating these Wilson coefficients is above the LHC scale, this constraint would become complementary to the one obtained from LHC observables. 
Analogously, if the $\nu$SMEFT is the valid EFT at the TeV energies, the potential $D$ parameter bounds on  $\tilde{\epsilon}_X$ translates into bounds on 
$C_{e \nu u d}^{(1)}$, $C_{e \nu u d}^{(3)}$, $C_{l\nu u q}$, and $C_{e\nu u d}$ in \cref{eq:EFT_scenario3,eq:EFT_scenario4}.

\begin{figure}[tb]
    \centering
    \includegraphics[width=0.5\textwidth]{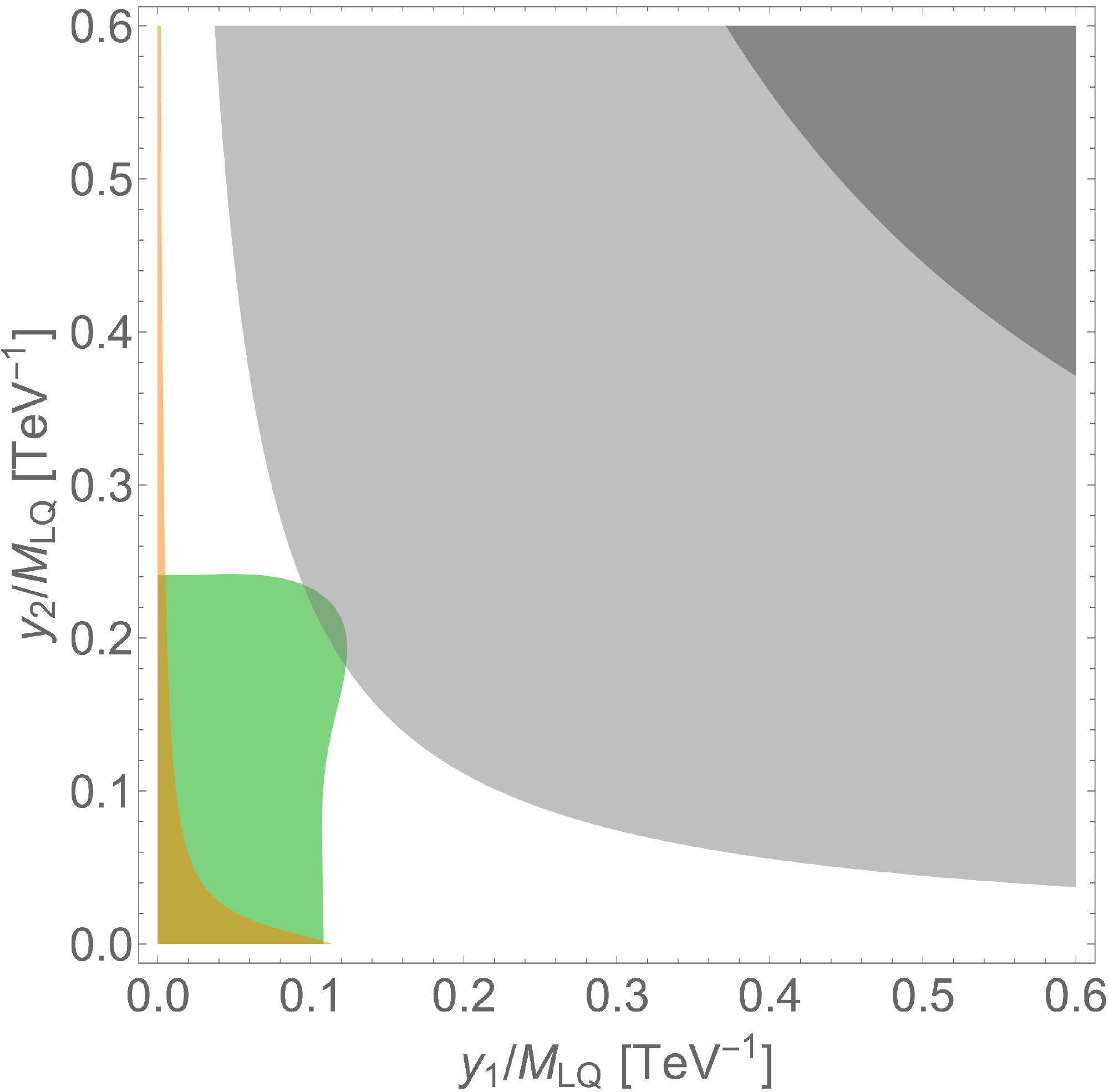}
    \caption{
    Constraints on the parameters of the $S_1$-$R_2$ leptoquark model with the cubic interactions in \cref{eq:LOOP_model}. 
    We show the region preferred at  95\% CL by the LHC $pp \to e^+ e^-$ Drell-Yan data~\cite{Allwicher:2022mcg}  (green), and by the combination of low-energy precision measurements performed in Ref.~\cite{Cirigliano:2021yto} (orange).  
We also show the region excluded by a hypothetical measurement of the $D$ parameter in ${}^{23}$Mg with the uncertainty of $10^{-4}$ (darker shade of gray) and $10^{-5}$ (lighter shade of gray).    }
    \label{fig:cpeven}
\end{figure}

To compare the sensitivity of the $D$ parameter and LHC searches in a concrete BSM scenario, we turn to a particular leptoquark model.   
We consider the following (somewhat contrived) model with the  $S_1$ and  $R_2$ with the common mass $M_{\rm LQ}$ and the cubic interactions
\begin{align}
\label{eq:LOOP_model}
\mathcal{L}&  \supset 
y_1\, (S_1 q + R_2 u^{c})\,\tilde{l}
+y_2\,(S_1 \bar u^c -  R_2 \bar{q})\,\bar{e}^c 
+ \mathrm{h.c.} \, , 
\end{align}
where $y_{1,2}$ are real.
The model is designed to only generate the tensor but no scalar or pseudoscalar charged current interactions at the high scale $M_{\rm LQ}$. 
Indeed, using Eq.~(\ref{eq:LQ_matching}) one finds that integrating out the leptoquarks leads to, among others, the SMEFT Wilson coefficients 
$C_{lequ}^{(1)} = C_{ledq} = 0$ and 
$C_{lequ}^{(3)}= - {y_1 y_2 \over 4 M_{\rm LQ}^2 }$.
Without taking into account RG running this would translate to 
$\epsilon_S= \epsilon_P= 0$ 
and 
$\epsilon_T = {y_1 y_2 v^2  \over 2 V_{ud} M_{\rm LQ}^2}$
in the EFT below the electroweak scale. 
RG running is however relevant, because it generates  $\epsilon_P$.   
The future bound on the $D$ parameter in ${}^{23}$Mg beta decays would be another probe of this CP-conserving scenario, in addition to the LHC and low-energy precision measurements.  
In \cref{fig:cpeven} we illustrate the sensitivity of the different probes. A measurement of the $D$ parameter in ${}^{23}$Mg with the uncertainty of $10^{-5}$ would provide a competitive probe of the parameters of the model compared to the current constraints from the LHC $pp \to e^+ e^-$ Drell-Yan data.
In this analysis we assume $M_{LQ} \gg 1$~TeV, such that the effects of leptoquark on Drell-Yan production can be correctly estimated  by dimension-6 operators in the $\nu$SMEFT. 
If the leptoquarks are lighter, say in the 2 TeV ballpark, then the Drell-Yan constraints become actually {\em weaker}, further increasing the relevance of the $D$ parameter measurements.
On the other hand, the low-energy precision measurements are in this specific scenario much  stronger than the LHC or the $D$ parameter.
This is because RG running generates non-zero $\epsilon_P$ at $\mu \simeq 2$~GeV, which is very strongly constrained by pion decay.\footnote{The main point of this exercise was to compare the LHC and $D$ parameter sensitivities in a concrete BSM setting. For the record, however, we note that one could avoid the pion constraint by a small adjustment of the leptoquark couplings in \cref{eq:LOOP_model} so that $\epsilon_P(2\, \mathrm{GeV})=0$ after running. Of course,  such a fine-tuning  seems difficult to  motivate from the UV perspective.}

\section{Conclusions} 

In this paper we discussed the prospects to uncover new physics beyond the SM via measurements of the $D$ correlation coefficient in beta decay.
The experimental sensitivity is expected to improve to $\Delta D \sim 10^{-5}$ in the coming decade~\cite{Delahaye:2018kwf}. 
This should allow one to measure the small $D \sim 10^{-4}$ predicted by the SM due final-state electromagnetic interactions between the daughter nucleus and the beta particle. 
More generally, $D$ probes imaginary parts of Wilson coefficients in the general nucleon-level EFT, which are currently not constrained by other experiments. 
Therefore, future measurements of the $D$ parameter will not only provide a test of the SM but also add unique information about the EFT for beta decay. 

A separate question is: which concrete models of new physics can be discovered or constrained by the $D$ parameter measurements? 
In principle, new heavy particle with CP-violating interactions may induce complex phases of EFT Wilson coefficients, and thus contribute to $D$.  
We argue however that realistic models leading to an observable shift from the SM prediction would have to be severely fine-tuned to avoid simultaneous constraints from other experiments, especially from  measurements of electric dipole moments, but also from pion decay and Drell-Yan electron pair production at the LHC.   
We give a very general, model-independent argument for this assertion, working  at the level of an EFT below and above the electroweak scale. 
We classify the scenarios for generating the $D$ parameters in this EFT and identify the problems with each scenario.   
The constraints and necessary tunings are also illustrated in concrete BSM settings involving leptoquarks. 
Our conclusions update, generalize,  and strengthen those of Ref.~\cite{Ng:2011ui} given the experimental progress in the last decade. 
Since the sensitivity of EDM experiments is expected to improve fast in the coming years, we expect our conclusions to only become stronger as time goes by. 

A word of caution is in order.  
It should be stressed that our results do not exclude in a model-independent way that a signal of CP violation appears in the next generation of experiments measuring the $D$ parameter. 
Non-trivial relations between parameters of BSM models may arise because of symmetries or by accident, leading to an apparent fine tuning from the low-energy perspective. 
If one allows for such fine tuning, CP-violating contributions to the $D$ parameter can be significant, even $\cO(10^{-4})$. 
What we claim is that, if a deviation from the SM prediction is found,  
the BSM models explaining it via new CP violating phases will necessarily appear fine-tuned or involve an element of baroqueness.

We also discussed the possibility of CP-conserving BSM contributions to the $D$ parameter, reaching more optimistic conclusions. 
While the $D$ correlation is T-odd,  
final state interactions at one loop level can provide necessary phases to induce $D$ even in the absence of fundamental CP violation. 
In particular, the real parts of non-standard scalar and tensor charged  currents in the nucleon-level EFT interfere with the electromagnetic Coulomb corrections to beta decay so as to generate $D$~\cite{Jackson:1957auh}. 
Even though the scalar and tensor current must be suppressed compared to the standard vector and axial currents~\cite{Falkowski:2020pma}, 
their contribution to $D$ arises at the leading order in $1/m_N$ expansion, unlike the SM one which arise at the next-to-leading order.  
For this reason, the $D$ parameter may soon become a sensitive probe of CP-conserving non-standard currents. 
We find that the experimental uncertainty of order $\Delta D \sim 10^{-5}$ will be enough to compete with other sensitive beta decay probes (superallowed and neutron decays), as well as with the LHC observables.

\section*{Acknowledgements}
We would like to thank Pierre~Delahaye and Olcyr~Sumensari for useful discussions. We are very grateful to the authors of Ref.~\cite{Allwicher:2022mcg} for making their code available to us before publication. AF and ARS are  supported by the Agence Nationale de la Recherche (ANR) under grant ANR-19-CE31-0012 (project MORA).
AF is supported by the European Union's Horizon 2020 research and innovation programme under the Marie Sklodowska-Curie grant agreement No. 860881 (HIDDe$\nu$ network). ARS was supported by European Union’s Horizon 2020 research and innovation programme under grant agreement No 101002846, ERC CoG ``CosmoChart".

\appendix

\section{Classification of leptoquarks} 
\label{sec:lepto}

In this appendix we review the classification and interactions of leptoquarks, as well as the matching between the parameters of leptoquark models and the $\nu$SMEFT effective operators  relevant for our analysis. 
Leptoquarks are bosons that can have an interaction vertex with a quark and a lepton.
In this paper we restrict to leptoquarks coupling only to the first generation of the SM fermions.  
For spin-0 and spin-1 leptoquarks, the representations under the SM gauge group that allow for such interactions have been catalogued~\cite{Buchmuller:1986zs,Dorsner:2016wpm} and reproduced here in \cref{tab:LQ_spin0}.
In this paper we forbid baryon number violating interactions, such that leptoquarks can be assigned baryon and lepton numbers that are conserved in perturbation theory.

\begin{table}[tb]
\bc 
{
\begin{tabular}{|c|c|c|c|c|}
\hline 
Name & Spin & Representation & Couplings 
\\ \hline
$S_1$ & 0 &  $({\bf \bar 3,1,1/3 })$ &  $q l$,  $\bar u^c \bar e^c$,  $\bar d^c \bar \nu^c$ 
\\ \hline
$\bar S_1$ &  0 & $({\bf \bar 3,1,-2/3 })$  &   $\bar u^c \bar \nu^c$ 
\\ \hline
$\tilde S_1$ & 0 &  $({\bf \bar 3,1,4/3 })$ &   $\bar d^c \bar e^c$ 
\\ \hline
$R_2$ &  0 &  $({\bf 3,2,7/6 })$ &    $u^c  l$, $\bar q \bar e^c$     
\\ \hline 
$\tilde R_2$ &  0 &  $({\bf 3,2,1/6 })$ &   $d^c  l $, $\bar q \bar \nu^c$ 
\\ \hline
$S_3$ &  0 &  $({\bf \bar 3,3,1/3 })$ &  $q \sigma^k l$  
\\ \hline \hline 
$U_1$ & 1 & $({\bf  3,1,2/3 })$ &  $\bar q \bar \sigma^\mu l$,  $d^c  \sigma^\mu \bar e^c$,  $u^c \sigma^\mu \bar \nu^c$ 
\\ \hline
$\bar U_1$ &  1 & $({\bf 3,1,-1/3 })$  &   $d^c  \sigma^\mu \bar \nu^c$ 
\\ \hline
$\tilde U_1$ &  1 &  $({\bf 3,1,5/3 })$ &   $u^c \sigma^\mu \bar e^c$ 
\\ \hline
$V_2$ & 1 &  $({\bf \bar 3,2,5/6 })$ &    $l \sigma^\mu \bar d^c $, $q \sigma^\mu \bar e^c$     
\\ \hline 
$\tilde V_2$ & 1 &  $({\bf \bar 3,2,-1/6 })$ &   $l \sigma^\mu \bar u^c $, $q \sigma^\mu \bar \nu^c$ 
\\ \hline
$U_3$ &  1 & $({\bf 3,3,2/3 })$ &  $\bar q \sigma^k  \bar \sigma^\mu l$  
\\ \hline
\end{tabular}
}
\ec 
\caption{
\label{tab:LQ_spin0}
Complete list of scalar and vector leptoquarks that can have tri-linear couplings to a quark and a lepton. The nomenclature follows the one used in Refs.~\cite{Buchmuller:1986zs,Dorsner:2016wpm}.
}
\end{table}

One can show that the $\bar S_1$, $\tilde S_1$, $S_3$, $\bar U_1$, $\tilde U_1$, and $U_3$  leptoquarks do not contribute to the $D$ parameter, therefore we will not consider them in the following. 
A simple argument is that the quantum numbers of these leptoquarks permit only a single trilinear interaction term with a quark and a lepton, in which case the phase of the coupling constant can be always eliminated by rephasing the leptoquark field. 
On the other hand, irreducible phases can be present in the interactions of  $S_1$, $R_2$, $\tilde R_2$,  $U_1$, $V_2$, and $\tilde V_2$ if they couple to at least two distinct lepton-quark structures.  
For these leptoquarks, we consider a general Lagrangian of the form 
\begin{equation}
\label{eq:LQ_L}
{\cal L}_{\rm leptoquark} = {\cal L}_{\rm kinetic} + {\cal L}_{\rm scalar} + {\cal L}_{vector} 
\end{equation}
Above, ${\cal L}_{\rm kinetic}$ contains the kinetic and mass terms of $S_1$, $R_2$, $\tilde R_2$,  $U_1$, $V_2$, and $\tilde V_2$.  
${\cal L}_{\rm scalar}$ contains Yukawa interactions for the scalar leptoquarks: 
\begin{align}
\label{eq:LQ_Lscalar}
{\cal L}_{\rm scalar} = &
 S_1 \Big [ y_{S q l } q  \tilde l  
+ y_{S u e}  \bar u^c \bar e^c 
+ y_{S d \nu}  \bar d^c \bar \nu^c 
\Big ] 
\nnl + & 
R_2 \Big [   y_{R u l}   u^c  \tilde l  
+ y_{R q e} \bar q  \bar e^c   
\Big ] 
+ \tilde R_2  \Big [ y_{R d l} d^c  \tilde l  
+ y_{R q \nu}  \bar q  \bar \nu^c  
\Big ] 
+ \hc        
\end{align}
${\cal L}_{\rm vector}$ contains all possible baryon-number conserving tri-linear interactions of vector leptoquark with a quark and a lepton:
\begin{align}
\label{eq:LQ_Lvector}
{\cal L}_{\rm vector} = &
U_1^\mu \Big [ 
g_{U q l }  \bar q \bar \sigma_\mu  l  
+ g_{U d e}  d^c  \sigma_\mu  \bar e^c 
+ g_{U u \nu} u^c \sigma_\mu \bar \nu^c 
\Big ] 
\nnl + & 
V_2^\mu \Big [   g_{Vdl}   \bar d^c  \bar \sigma_\mu  \tilde l  
+ g_{Vqe}    \tilde q \sigma_\mu  \bar e^c   
\Big ] 
+  \tilde V_2^\mu  \Big [ 
g_{ V u l}  \bar u^c  \bar \sigma_\mu  \tilde l  
+ g_{V q \nu} \tilde q  \sigma_\mu  \bar \nu^c  
\Big ] 
+ \hc   
\end{align}

Integrating out the leptoquarks, we obtain the following Wilson coefficients of the $\nu$SMEFT operators defined in \cref{eq:EFT_scenario2,eq:EFT_scenario3,eq:EFT_scenario4}:
\begin{align}
\label{eq:LQ_matching}
 C_{lequ}^{(1)} = & 
 {y_{S u e} \bar y_{S q l }  \over 2 M_{S_1}^2 }  
 +   { y_{R q e}  \bar  y_{R u l} \over 2 M_{R_2}^2}  , 
\nnl 
 C_{lequ}^{(3)} = &  
 - { y_{S u e} \bar y_{S q l } \over 8 M_{S_1}^2}  
 +  { y_{R q e}  \bar  y_{R u l} \over 8 M_{R_2}^2}  , 
 \nnl 
C_{ledq} = & 
2 {g_{U d e}  \bar g_{U q l } \over M_{U_1}^2}  
+ 2 {g_{Vqe}   \bar g_{Vdl} \over M_{V_2}^2} , 
 \nnl 
   C_{l \nu q  d}^{(1)}  = &  
   {y_{S d \nu} \bar y_{S q l }   \over 2 M_{S_1}^2}  
   +   {y_{R q \nu} \bar y_{R d l}       \over 2 M_{\tilde R_2}^2}  , 
   \nnl 
  C_{l \nu q  d}^{(3)}  = &   
  - { y_{S d \nu} \bar y_{S q l }  \over 8 M_{S_1}^2}    
  +    {  y_{R q \nu}  \bar y_{R d l}     \over 8 M_{\tilde R_2}^2}  , 
\nnl 
C_{l\nu u q} = & 
2  { g_{U u \nu}  \bar g_{U q l }\over M_{U_1}^2}
+ 2 {g_{V q \nu}  \bar g_{ V u l}  \over M_{\tilde V_2}^2}  , 
  \nnl 
  C_{e \nu u d} = &   
  { y_{S d \nu} \bar y_{S u e}   \over 2 M_{S_1}^2} 
  - {  g_{U u \nu} \bar g_{U d e}   \over M_{U_1}^2} . 
\end{align}  
As expected, the Wilson coefficients pick up an imaginary part when there is a relative phase between two different tri-linear couplings of the same leptoquark. 
That phase is a necessary but not sufficient condition for generating the D parameter.
As is clear from \cref{eq:EFT_Dscenario2,eq:EFT_Dscenario3}, to this end we also need a relative phase between the tensor Wilson coefficient ($C_{lequ}^{(3)}$ or $C_{l \nu q  d}^{(3)} $) and the scalar ones ($ C_{lequ}^{(1)}$, $C_{ledq}$, $   C_{l \nu q  d}^{(1)}$,  $C_{l\nu u q}$). 
Inspection of \cref{eq:LQ_matching} shows that the relative phase can arise  only if two distinct leptoquarks with different quantum numbers contribute. 
In other words, a (purely) leptoquark scenario for generating an observable D parameter must contain at least two leptoquarks near the TeV scale coupled to the first generation fermions.  
Finally, we note that leptoquarks do not generate the Wilson coefficients $C_{\phi u d}$ and  $C_{\phi e \nu}$ in \cref{eq:EFT_Dscenario1a,eq:EFT_Dscenario4}. 
Therefore scenarios Ia and IV in our nomenclature cannot arise from purely leptoquark BSM models at tree level.  
They can however arise in hybrid models, for example when a leptoquark is accompanied by an exotic right-handed W' vector boson. 

So far we haven't discussed how to connect leptoquarks to the scenario Ib where the D parameter is generated through the dimension-8 $\nu$SMEFT operator in \cref{eq:EFT_scenario1b}. 
That operator is not induced by integrating out the leptoquarks from the Lagrangian in \cref{eq:LQ_L}. 
However, there is a natural generalization of \cref{eq:LQ_L} to include cubic and quartic interactions of leptoquarks with the Higgs field.
Such interactions result in mixing between distinct leptoquarks after electroweak symmetry breaking, and lead to additional operators involving the Higgs field in the $\nu$SMEFT effective theory below the leptoquark mass scale. 
All renormalizable interactions mixing two scalar leptoquarks were catalogued in Ref.~\cite{Dorsner:2019itg}. 
Among those, the quartic interaction
\begin{equation}
\label{eq:LQ_RRmixing}
\Delta {\cal L}_{\rm leptoquark} = 
\lambda_{RR}  (R_2^\dagger H) (\tilde H^\dagger \tilde R_2  )     + \hc 
\end{equation}
induces the dimension-8 operator in \cref{eq:EFT_scenario1b} with the Wilson coefficient
\begin{equation}
C_8 = - {\lambda_{RR} y_{Rul} \bar y_{R d l} \over 2 M_{R_2}^2 M_{\tilde R_2}^2}.    
\end{equation}
One can show that the remaining mixing interactions discussed in \cite{Dorsner:2019itg} induce dimension-6 and 8 operators that are not relevant for the D parameter, and therefore they are not considered in this paper. 
In analogy to \cref{eq:LQ_RRmixing} on can also consider mixing of the vector leptoquarks via the (non-renormalizable) quartic interaction
\begin{equation}
\label{eq:LQ_TTmixing}
\Delta {\cal L}_{\rm leptoquark} = 
\lambda_{VV}  (V_2^\dagger H) (\tilde H^\dagger \tilde V_2  )     + \hc 
\end{equation}
This also induces the  dimension-8 operator in \cref{eq:EFT_scenario1b} with the Wilson coefficient
\begin{equation}
C_8 = - {\lambda_{VV} g_{Vdl} \bar g_{V u l} \over M_{V_2}^2 M_{\tilde V_2}^2}.     
\end{equation}
All in all, scenario Ib for generating the D parameter can be realized in a BSM model with two distinct SU(2) doublet leptoquarks that mix after electroweak symmetry breaking. 
Such models were already considered in Ref.~\cite{Ng:2011ui}, and in this paper we will update that analysis taking into account new experimental constraints, in particular from the LHC. 

\vspace{1cm}

We close this appendix with a discussion of other dimension-6 $\nu$SMEFT operators generated at tree level by integrating out the leptoquarks from the Lagrangian \cref{eq:LQ_L}. 
In addition to the operators defined in \cref{eq:EFT_scenario2,eq:EFT_scenario3,eq:EFT_scenario4}, which can have complex Wilson coefficients, there arise certain 4-fermion operators with real Wilson coefficients. 
The latter are irrelevant for the sake of CP violation, but they may be important  for CP-conserving precision observables in low-energy experiments and at the LHC. 
The set of dimension-6 CP-conserving dimension-6 operators generated by integrating out leptoquarks is 
\begin{align}
\label{eq:LQ_LeffCPconserving}
{\cal L}_{\nu \rm SMEFT } \supset  &
C_{lq}^{(1)} (\bar l \bar \sigma^\mu l) (\bar q \bar \sigma^\mu q) 
+ C_{lq}^{(3)}  (\bar l \bar \sigma^\mu \sigma^k l)  (\bar q \bar \sigma^\mu  \sigma^k q)
+ C_{eu}  (e^c \sigma^\mu \bar e^c ) (u^c  \sigma^\mu \bar u^c ) 
+ C_{ed}  (e^c \sigma^\mu \bar e^c ) (d^c  \sigma^\mu \bar d^c ) 
\nnl  + & 
 C_{lu}  (\bar l \bar \sigma^\mu l)  (u^c  \sigma^\mu \bar u^c ) 
+ C_{ld}  (\bar l \bar \sigma^\mu l)  (d^c  \sigma^\mu \bar d^c )
+ C_{qe}   (\bar q \bar \sigma^\mu q)  (e^c \sigma^\mu \bar e^c )
\nnl + & 
 C_{q\nu}   (\bar q \bar \sigma^\mu q)  (\nu^c \sigma^\mu \bar \nu^c )
+ C_{\nu u}  (\nu^c \sigma^\mu \bar \nu^c ) (u^c  \sigma^\mu \bar u^c ) 
+ C_{\nu d}  (\nu^c \sigma^\mu \bar \nu^c ) (d^c  \sigma^\mu \bar d^c ) . 
\end{align}
Matching the Wilson coefficients at tree level one finds 
\begin{align}
 C_{lq}^{(1)} = & { | y_{S q l }|^2  \over 4  M_{S_1}^2}  - {|g_{U q l }|^2 \over 2 M_{U_1}^2} , 
  \nnl 
 C_{lq}^{(3)}  = & - { | y_{S q l }|^2  \over 4  M_{S_1}^2}  - {|g_{U q l }|^2 \over 2 M_{U_1}^2}  , 
 \nnl  
 C_{eu}  = & {| y_{S u e}  |^2  \over 2  M_{S_1}^2} ,  
 \nnl 
 C_{ed} = & -{ |g_{U d e}|^2  \over M_{U_1}^2}  , 
 \nnl 
   C_{lu} = &   - {  |y_{R u l}|^2 \over 2 M_{R_2}^2} +    { | g_{ V u l} |^2 \over M_{\tilde V_2}^2} , 
 \nnl 
  C_{ld} = & -    { |y_{R d l}|^2   \over 2 M_{\tilde R_2}^2 }   +   { |g_{Vdl}|^2  \over M_{V_2}^2} ,   
 \nnl 
 C_{qe}   = &  - {|y_{R q e}|^2  \over 2 M_{R_2}^2}  +  { |g_{Vqe} |^2   \over M_{V_2}^2}  ,
  \nnl
  C_{q\nu} = & - { |y_{R q \nu}|^2 \over 2 M_{\tilde R_2}^2 } +  {  |g_{V q \nu}|^2  \over M_{\tilde V_2}^2} , 
    \nnl
 C_{\nu u}  = & -{| g_{U u \nu}|^2   \over M_{U_1}^2}  , 
 \nnl
 C_{\nu d}  = & { |y_{S d \nu} |^2 \over  2  M_{S_1}^2} .  \label{eq:LQwilsons}
\end{align}

\section{Coulomb corrections} 
\setcounter{equation}{0}   
\label{sec:dloops}  

In this appendix we present a concise derivation of the Coulomb corrections to the correlation coefficients in beta decay, originally calculated by  Jackson, Treiman, and  Wyld in Ref.~\cite{Jackson:1957auh}. 
As a particular application, we reproduce the Coulomb corrections to the $D$ parameter in \cref{eq:LOOP_Dnucleon}. 
For simplicity, in the derivation we only consider $\beta^-$ decay with the parent and daughter nuclei having the same spin $J$, and ignore interactions involving  the right-handed neutrino (effectively setting the $C_X^-$ Wilson coefficients to zero). A more general derivation would proceed along the same lines.   

The leading order amplitude for $\beta^-$ same-spin decay can be written in the following form (see~\cite{Falkowski:2021vdg}) 
\begin{align}
\label{eq:LOOP_M0}
\cM^{(0)} ({\cal N}_1 \to {\cal N}_2 e^- \bar \nu) =   2 m_{\cal N} M_F   \bigg \{  
-    \delta_{J_2^z }^{\, J_1^z}   \big [  C_V^+ L^0 + C_S^+ L \big ]  
+    { r\over \sqrt{J(J+1)} }  [{\cal T}_{(J)}^{k}]_{J_2^z}^{\, J_1^z}  \big [  C_A^+ L^k + C_T^+ L^{0k}  \big ] 
\bigg \} .  
\end{align}
Above, $C_X^+$ are the Wilson coefficients in the EFT Lagrangian of \cref{eq:EFT_pionlessL0}. 
$J_1^z$ and $J_2^z$ label the polarizations of the parent and daughter nuclei. 
The leptonic currents are defined as 
$L^\mu \equiv \bar x_3 \bar \sigma^\mu y_4$,
$L \equiv  y_3 y_4$, 
$L^{0k} \equiv y_3  \sigma^0 \bar \sigma^k y_4$,
where $k=1\dots 3$,  and $x_3$, $y_3$ are the 2-component spinor wave functions of the outgoing electron 
and $y_4$ is the spinor wave functions of the outgoing antineutrino.\footnote{%
In the 4-component Dirac formalism $u_3 = (x_3,\bar y_3)^T$, 
 $\bar u_3 = (y_3,\bar x_3)$, $v_4 = (y_4, \bar x_4)^T$.  } 
Next, ${\cal T}_{(J)}^k$ are the spin-J generators of the rotation group. 
We write down the amplitude in the limit of unbroken isospin symmetry, thus $m_{{\cal N}_2} = m_{{\cal N}_1} \equiv m_{\cal N}$. 
The common normalization factor is
$M_F =   \delta_{j_3',j_3 + 1}  \sqrt{ j (j+1) - j_3( j_3 + 1)}$, 
where $(j,j_3)$ and $(j,j_3')$ are the isospin quantum numbers of the parent and daughter nuclei. 
The parameter $r$, which is real by time-reversal invariance,  is referred to as the ratio of Gamow-Teller and Fermi matrix elements in the literature. 
For the neutron decay $r=\sqrt{3}$, while for nuclear beta decay $r$ is extracted from experiment. 

Next, we need the 1-loop corrections to this amplitude due to a photon exchange between the daughter nucleus and the electron. 
We denote this amplitude by 
$\cM^{(1)} ({\cal N}_1 \to {\cal N}_2 e^- \bar \nu)$.  
Actually, we only need the imaginary part of $\cM^{(1)}$, which is determined by unitarity using the master formula
 \begin{align}
 \label{eq:LOOP_master}
\im \cM^{(1)} ({\cal N}_1 \to {\cal N}_2 e^- \bar \nu) = {1 \over 2} \sum_{h_{3'} J_{2'}^z } \int d \Pi'_2 \cM^{(0)} ({\cal N}_1 \to {\cal N}_2 e^- \bar \nu) 
  \cM_{\rm em}  ( {\cal N}_2 e^- \to  {\cal N}_2 e^-) . 
 \end{align}
Here, $d \Pi'_2$ denotes the 2-body phase space of the intermediate $ {\cal N}_2 e^-$ pair, 
$ J_{2'}^z$ is the polarization of the intermediate nucleus, and $h_{3'}$ is he polarization of the intermediate electron. 
$\cM_{\rm em}$ is the tree-level scattering amplitude due to a photon exchange   between the daughter nucleus and the electron. 
Expanding it in $1/m_{\cal N}$, for an arbitrary spin $J$ the leading piece takes the form 
 \begin{align}
 \label{eq:LOOP_Mem}
  \cM_{\rm em}  ( {\cal N}_2 e^- \to  {\cal N}_2 e^-)  = & 
 \delta_{J_{2}^z }^{\, J_{2'}^z}    {2 q_e Z e^2 \over (k_{3} - p_{3'} )^2 }    \big [ \bar x_{3} p_{2'} \bar \sigma x_{3'}  + y_{3} p_{2'} \sigma \bar y_{3'} \big ]  . 
 \end{align}
 Above, $q_e = -1$, $Z$ is a (positive) charge of the daughter nucleus, $e \sim 0.3$ is the electromagnetic coupling constant, 
the incoming momenta of the nucleus and electron are denoted by $p_{2'}$ and $p_{3'}$, 
and the outgoing  momentum of the electron is denoted by  $k_{3}$.  
The 2-component spinor wave function of the incoming electron are denoted as $x_{3'}$ and $y_{3'}$, 
and  the spinor wave function of the outgoing electron are denoted as $x_{3}$ and $y_{3}$. 

Plugging \cref{eq:LOOP_Mem,eq:LOOP_M0} into \cref{eq:LOOP_master} we get
 \begin{align}
 \label{eq:LOOP_MbeforeSpinSum}
\im \cM^{(1)} ({\cal N}_1 \to {\cal N}_2 e^- \bar \nu) =  &  2 m_{\cal N} M_F  q_e Z e^2    \sum_{h_{3'} } \int d \Pi'_2 
   {1 \over (k_{3} - p_{3'} )^2 }    \big [ \bar x_{3} p_{2'} \bar \sigma x_{3'}  + y_{3} p_{2'} \sigma \bar y_{3'} \big ]   
   \nnl & 
     \bigg \{    
-    \delta_{J_2^z }^{\, J_1^z}   \big [  C_V^+ L'{}^0 + C_S^+ L' \big ]  
+   { r\over \sqrt{J(J+1)} }  [{\cal T}_{(J)}^{k}]_{J_2^z}^{\, J_1^z}  \big [  C_A^+ L'{}^k + C_T^+ L'{}^{0k}  \big ] 
\bigg \} , 
 \end{align}
 where now $L'{}^\mu \equiv \bar x_{3'} \bar \sigma^\mu y_4$,
$L' \equiv  y_{3'} y_4$, 
$L'{}^{0k} \equiv y_{3'}  \sigma^0 \bar \sigma^k y_4$. 
We perform the $h_{3'}$ sum over the intermediate electron polarizations to derive  
\begin{align} 
\label{eq:LOOP_spinsums}
  \sum_{h_{3'} } \big [ \bar x_{3} p_{2'} \bar \sigma x_{3'}  + y_{3} p_{2'} \sigma \bar y_{3'} \big ]   L'{}^\mu 
  = &  (k_2 k_3) L^\mu  + m_e p_{2'}^\mu L  + m_e  p_{2'}{}_\nu L^{\nu \mu} +  p_{3'}^\mu  p_{2'}{}_\nu L^\nu -   p_{2'}^\mu  p_{3'}{}_\nu L^\nu
  \nnl & 
  - i \epsilon^{\alpha \beta \mu \nu} p_{2'}{}_\alpha  p_{3'}{}_\beta L_\nu, 
  \nnl 
    \sum_{h_{3'} } \big [ \bar x_{3} p_{2'} \bar \sigma x_{3'}  + y_{3} p_{2'} \sigma \bar y_{3'} \big ]   L' = & 
      (k_2 k_3) L  + m_e p_{2'}{}_\mu L^{\mu}  + p_{2'}{}_\mu  p_{3'}{}_\nu L^{\mu\nu}, 
        \nnl 
\sum_{h_{3'} } \big [ \bar x_{3} p_{2'} \bar \sigma x_{3'}  + y_{3} p_{2'} \sigma \bar y_{3'} \big ]   L'{}^{0k} = &  
(k_2 k_3) L^{0k}  + m_e p_{2'}^0 L^k -  m_e p_{2'}^k L^0 +    p_{2'}^k p_{3'}^l L^{0l} -  p_{3'}^k p_{2'}^l L^{0l} 
\nnl  & 
+  p_{2'}^k p_{3'}^0 L  - p_{3'}^k p_{2'}^0 L
- i m_e \epsilon^{klm}   p_{2'}^l L^m 
- i \epsilon^{kl \alpha \beta}  p_{2'}{}_\alpha  p_{3'}{}_\beta  L^{0l}
+ i \epsilon^{klm}p_{2'}^l p_{3'}^m . 
 \end{align}
 We insert the spin sums \cref{eq:LOOP_spinsums} into \cref{eq:LOOP_MbeforeSpinSum}. 
 Working at the leading order in $1/m_{\cal N}$ we can replace in the spin sums 
 $p_{2'}^0 \to m_{\cal N}$, $p_{2'}^k \to 0$,  $k_2 k_3 \to m_{\cal N} E_e$. 
 With these replacement, the spin-summed \cref{eq:LOOP_MbeforeSpinSum} becomes 
  \begin{align}
 \label{eq:LOOP_MafterSpinSum}
\im \cM^{(1)} ({\cal N}_1 \to {\cal N}_2 e^- \bar \nu) =  &  2 m_{\cal N}^2 M_F  q_e Z e^2  \int d \Pi'_2 
   {1 \over (k_{3} - p_{3'} )^2 }       \bigg \{ 
      \nnl & 
-   \delta_{J_2^z }^{\, J_1^z} C_V^+   \big [ E_e L^0 + m_e L + p_{3'}^k L^k  \big ] 
-   \delta_{J_2^z }^{\, J_1^z}  C_S^+ \big [  E_e L + m_e L^0  - p_{3'}^k L^{0k}  \big ]  
      \nnl & 
+   { r\over \sqrt{J(J+1)} }  [{\cal T}_{(J)}^{k}]_{J_2^z}^{\, J_1^z}    C_A^+ \big [  
 E_e L^k + m_e L^{0k} + p_{3'}^k L^0 +  i \epsilon^{klm}  p_{3'}^l L^m  \big ]
      \nnl & 
+   { r\over \sqrt{J(J+1)} }  [{\cal T}_{(J)}^{k}]_{J_2^z}^{\, J_1^z} C_T^+   \big [ 
E_e L^{0k}+ m_e L^{k} -  p_{3'}^k L -    i \epsilon^{klm}  p_{3'}^l L^{0m} \big ] 
\bigg \} . 
 \end{align}
We can now integrate over the intermediate phase space using 
   \begin{align}
  \int d \Pi'_2    {1 \over (k_{3} - p_{3'} )^2 }      =&  
  {1 \over 16 \pi p_e m_{\cal N} } \int_{-1}^1 d \cos \theta {1  \over  \cos \theta - 1} ,   
  \nnl 
    \int d \Pi'_2    {p_{3'}^k \over (k_{3} - p_{3'} )^2 }      =&  
  {k_{3}^k  \over 16 \pi p_e m_{\cal N} } \int_{-1}^1  d \cos \theta {\cos \theta  \over \cos \theta - 1} , 
    \end{align}
    where for the time being we ignore the IR divergence corresponding to $\cos \theta \to 1$ (forward re-scattering). 
   This plus some more spinor algebra leads to 
     \begin{align}
 \label{eq:LOOP_Mcoulomb}
\im \cM^{(1)} ({\cal N}_1 \to {\cal N}_2 e^- \bar \nu) =  &  2 m_{\cal N} M_F  
{(-q_e)  Z \alpha  \over  4 p_e}    \int_{-1}^1 d \cos \theta     \bigg \{ 
-   \delta_{J_2^z }^{\, J_1^z} C_V^+   \big [  E_e L^0  {1 + \cos \theta   \over 1 - \cos \theta }  +  m_e L   \big ] 
  \nnl - &  
  \delta_{J_2^z }^{\, J_1^z}  C_S^+ \big [  E_e L  {1 + \cos \theta   \over 1 - \cos \theta }  +  m_e L^0    \big ]  
+ { r\over \sqrt{J(J+1)} }  [{\cal T}_{(J)}^{k}]_{J_2^z}^{\, J_1^z}    C_A^+ \big [
  E_e L^k  {1 + \cos \theta   \over 1 - \cos \theta }   + m_e L^{0k}   \big ]
      \nnl + & 
 { r\over \sqrt{J(J+1)} }  [{\cal T}_{(J)}^{k}]_{J_2^z}^{\, J_1^z} C_T^+   \big [ 
 E_e L^{0k} {1 + \cos \theta   \over 1 - \cos \theta }  +  m_e L^{k} \big ] 
\bigg \}.  
 \end{align}
Combine now the Coulomb correction with the leading order amplitude in \cref{eq:LOOP_M0} we obtain $\cM \equiv \cM^{(0)} + i \im \cM^{(0)}$: 
     \begin{align}
 \label{eq:LOOP_Mcombined}
\cM ({\cal N}_1 \to {\cal N}_2 e^- \bar \nu) =  &  2 m_{\cal N} M_F  
    \bigg \{ 
-   \delta_{J_2^z }^{\, J_1^z} C_V^+   \big [ L^0 (1 + {\color{blue} i x} )   
+  {\color{red} i {\Delta_Z \over 2}  L}  \big ] 
      \nnl - & 
   \delta_{J_2^z }^{\, J_1^z}  C_S^+ \big [  L (1 +  {\color{blue} i x } )  + {\color{red} i  {\Delta_Z \over 2}   L^0}   \big ]  
+   { r\over \sqrt{J(J+1)} }  [{\cal T}_{(J)}^{k}]_{J_2^z}^{\, J_1^z}    C_A^+ \big [
L^k (1 +  {\color{blue} i x })  
+  {\color{red} i  {\Delta_Z \over 2}  L^{0k}}   \big ]
      \nnl + & 
   { r\over \sqrt{J(J+1)} }  [{\cal T}_{(J)}^{k}]_{J_2^z}^{\, J_1^z} C_T^+   \big [ 
L^{0k}  (1 +  {\color{blue} i x} ) 
+ {\color{red} i  {\Delta_Z \over 2}   L^{k}}  \big ] 
\bigg \} , 
 \end{align}
 where we  defined 
 \begin{align}
x \equiv   &  { (- q_e) Z \alpha E_e  \over  4 p_e}    
\int_{-1}^1 d \cos \theta {1 + \cos \theta  \over 1 - \cos \theta} , 
\qquad 
\Delta_Z \equiv    { (- q_e) Z \alpha m_e \over   p_e}   . 
 \end{align}
The effect of the blue terms in \cref{eq:LOOP_Mcombined} is an overall rescaling of all leading order Wilson coefficients by the same (IR divergent) phase factor: 
 $C_X^+  \to C_X^+ e^{i x}$. 
This does not change the correlation coefficients, 
as they always depend on the $C_X \bar C_Y$  combinations.  
On the other hand, the effect of the red terms is to rotate the leading order Wilson coefficients among each other, which does affect the correlations. 
All in all, the Coulomb corrections to the beta decay amplitude can be concisely described by the transformation  
\begin{align}
\label{eq:LOOP_zxRotation}
 C_V^\pm  \to  & C_V^\pm e^{\pm i x}  \pm  {i  \Delta_Z  \over 2}   C_S^\pm ,
 \qquad 
 C_S^\pm  \to   C_S^\pm e^{\pm i x}  \pm  {i \Delta_Z  \over 2}    C_V^\pm , 
 \nnl
  C_A^\pm \to  & C_A^\pm e^{\pm i x} \pm  {i \Delta_Z  \over 2}   C_T^\pm , 
\qquad  
 C_T^\pm \to  C_T^\pm e^{\pm i x}   \pm  {i \Delta_Z  \over 2}   C_A^\pm , 
\end{align}
where at this point we generalize the result to $\beta^\mp$ decays (the $\pm$ signs on the right-hand side), and to include the right-handed neutrino interactions (the $C_X^-$ Wilson coefficients). 
One can check that inserting this transformation in the leading order expression for the correlation coefficients (and ignoring $\cO(Z^2 \alpha^2)$ terms) reproduces the Coulomb corrections listed in Ref.~\cite{Jackson:1957auh}. 
One particular application of \cref{eq:LOOP_zxRotation} is to easily obtain the Coulomb corrections to the $D$ parameter in \cref{eq:LOOP_Dnucleon} at the $1/m_N^0$ order in the EFT.

\bibliographystyle{JHEP}
\bibliography{leptomoraV2}

\providecommand{\href}[2]{#2}\begingroup\raggedright\begin{thebibliography}{10}

\bibitem{Jackson:1957zz}
J.D.~Jackson, S.B.~Treiman and H.W.~Wyld, \emph{{Possible tests of time
  reversal invariance in Beta decay}},
  \href{https://doi.org/10.1103/PhysRev.106.517}{\emph{Phys. Rev.} {\bfseries
  106} (1957) 517}.

\bibitem{Callan:1967zz}
C.G.~Callan and S.B.~Treiman, \emph{{Electromagnetic Simulation of T Violation
  in Beta Decay}}, \href{https://doi.org/10.1103/PhysRev.162.1494}{\emph{Phys.
  Rev.} {\bfseries 162} (1967) 1494}.

\bibitem{Delahaye:2018kwf}
P.~Delahaye et~al., \emph{{The MORA project}},  in \emph{{7th International
  Conference on Trapped Charged Particles and Fundamental Physics (TCP 2018)
  Takamatsu, Japan, September 30-October 5, 2018}}, 2018
  [\href{https://arxiv.org/abs/1812.02970}{{\ttfamily 1812.02970}}].

\bibitem{Herczeg:2001vk}
P.~Herczeg, \emph{{Beta decay beyond the standard model}},
  \href{https://doi.org/10.1016/S0146-6410(01)00149-1}{\emph{Prog. Part. Nucl.
  Phys.} {\bfseries 46} (2001) 413}.

\bibitem{Ng:2011ui}
J.~Ng and S.~Tulin, \emph{{D versus d: CP Violation in Beta Decay and Electric
  Dipole Moments}},
  \href{https://doi.org/10.1103/PhysRevD.85.033001}{\emph{Phys. Rev.}
  {\bfseries D85} (2012) 033001}
  [\href{https://arxiv.org/abs/1111.0649}{{\ttfamily 1111.0649}}].

\bibitem{El-Menoufi:2016cfo}
B.K.~El-Menoufi, M.J.~Ramsey-Musolf and C.-Y.~Seng, \emph{{Right-Handed
  Neutrinos and T-Violating, P-Conserving Interactions}},
  \href{https://doi.org/10.1016/j.physletb.2016.11.061}{\emph{Phys. Lett. B}
  {\bfseries 765} (2017) 62}
  [\href{https://arxiv.org/abs/1605.09060}{{\ttfamily 1605.09060}}].

\bibitem{Ramsey-Musolf:2020ndm}
M.J.~Ramsey-Musolf and J.C.~Vasquez, \emph{{Left-right symmetry and electric
  dipole moments. A global analysis}},
  \href{https://doi.org/10.1016/j.physletb.2021.136136}{\emph{Phys. Lett. B}
  {\bfseries 815} (2021) 136136}
  [\href{https://arxiv.org/abs/2012.02799}{{\ttfamily 2012.02799}}].

\bibitem{Jackson:1957auh}
J.D.~Jackson, S.B.~Treiman and H.W.~Wyld, \emph{{Coulomb corrections in allowed
  beta transitions}},
  \href{https://doi.org/10.1016/0029-5582(87)90019-8}{\emph{Nucl. Phys.}
  {\bfseries 4} (1957) 206}.

\bibitem{Gonzalez-Alonso:2018omy}
M.~Gonzalez-Alonso, O.~Naviliat-Cuncic and N.~Severijns, \emph{{New physics
  searches in nuclear and neutron $\beta$ decay}},
  \href{https://doi.org/10.1016/j.ppnp.2018.08.002}{\emph{Prog. Part. Nucl.
  Phys.} {\bfseries 104} (2019) 165}
  [\href{https://arxiv.org/abs/1803.08732}{{\ttfamily 1803.08732}}].

\bibitem{Falkowski:2020pma}
A.~Falkowski, M.~Gonz\'alez-Alonso and O.~Naviliat-Cuncic, \emph{{Comprehensive
  analysis of beta decays within and beyond the Standard Model}},
  \href{https://doi.org/10.1007/JHEP04(2021)126}{\emph{JHEP} {\bfseries 04}
  (2021) 126} [\href{https://arxiv.org/abs/2010.13797}{{\ttfamily
  2010.13797}}].

\bibitem{Zyla:2020zbs}
{\scshape Particle Data Group} collaboration, \emph{{Review of Particle
  Physics}}, \href{https://doi.org/10.1093/ptep/ptaa104}{\emph{PTEP} {\bfseries
  2020} (2020) 083C01}.

\bibitem{Pierre}
P.~Delahaye, \emph{{The MORA experiment}}, {\emph{Presentation at ISOL@MYRRHA}
  (2022) }.

\bibitem{Falkowski:2021vdg}
A.~Falkowski, M.~Gonz\'alez-Alonso, A.~Palavri\'c and
  A.~Rodr\'\i{}guez-S\'anchez, \emph{{Constraints on subleading interactions in
  beta decay Lagrangian}},  \href{https://arxiv.org/abs/2112.07688}{{\ttfamily
  2112.07688}}.

\bibitem{vanKolck:1999mw}
U.~van Kolck, \emph{{Effective field theory of nuclear forces}},
  \href{https://doi.org/10.1016/S0146-6410(99)00097-6}{\emph{Prog. Part. Nucl.
  Phys.} {\bfseries 43} (1999) 337}
  [\href{https://arxiv.org/abs/nucl-th/9902015}{{\ttfamily nucl-th/9902015}}].

\bibitem{Dreiner:2008tw}
H.K.~Dreiner, H.E.~Haber and S.P.~Martin, \emph{{Two-component spinor
  techniques and Feynman rules for quantum field theory and supersymmetry}},
  \href{https://doi.org/10.1016/j.physrep.2010.05.002}{\emph{Phys.Rept.}
  {\bfseries 494} (2010) 1} [\href{https://arxiv.org/abs/0812.1594}{{\ttfamily
  0812.1594}}].

\bibitem{Lee:1956qn}
T.D.~Lee and C.-N.~Yang, \emph{{Question of Parity Conservation in Weak
  Interactions}}, \href{https://doi.org/10.1103/PhysRev.104.254}{\emph{Phys.
  Rev.} {\bfseries 104} (1956) 254}.

\bibitem{Ademollo:1964sr}
M.~Ademollo and R.~Gatto, \emph{{Nonrenormalization Theorem for the Strangeness
  Violating Vector Currents}},
  \href{https://doi.org/10.1103/PhysRevLett.13.264}{\emph{Phys. Rev. Lett.}
  {\bfseries 13} (1964) 264}.

\bibitem{Aoki:2021kgd}
Y.~Aoki et~al., \emph{{FLAG Review 2021}},
  \href{https://arxiv.org/abs/2111.09849}{{\ttfamily 2111.09849}}.

\bibitem{Gupta:2018qil}
R.~Gupta, Y.-C.~Jang, B.~Yoon, H.-W.~Lin, V.~Cirigliano and T.~Bhattacharya,
  \emph{{Isovector Charges of the Nucleon from 2+1+1-flavor Lattice QCD}},
  \href{https://doi.org/10.1103/PhysRevD.98.034503}{\emph{Phys. Rev.}
  {\bfseries D98} (2018) 034503}
  [\href{https://arxiv.org/abs/1806.09006}{{\ttfamily 1806.09006}}].

\bibitem{Chang:2018uxx}
C.~Chang et~al., \emph{{A per-cent-level determination of the nucleon axial
  coupling from quantum chromodynamics}},
  \href{https://doi.org/10.1038/s41586-018-0161-8}{\emph{Nature} {\bfseries
  558} (2018) 91} [\href{https://arxiv.org/abs/1805.12130}{{\ttfamily
  1805.12130}}].

\bibitem{Walker-Loud:2019cif}
A.~Walker-Loud et~al., \emph{{Lattice QCD Determination of $g_A$}},
  \href{https://doi.org/10.22323/1.317.0020}{\emph{PoS} {\bfseries CD2018}
  (2020) 020} [\href{https://arxiv.org/abs/1912.08321}{{\ttfamily
  1912.08321}}].

\bibitem{Liao:2016qyd}
Y.~Liao and X.-D.~Ma, \emph{{Operators up to Dimension Seven in Standard Model
  Effective Field Theory Extended with Sterile Neutrinos}},
  \href{https://doi.org/10.1103/PhysRevD.96.015012}{\emph{Phys. Rev. D}
  {\bfseries 96} (2017) 015012}
  [\href{https://arxiv.org/abs/1612.04527}{{\ttfamily 1612.04527}}].

\bibitem{Li:2021tsq}
H.-L.~Li, Z.~Ren, M.-L.~Xiao, J.-H.~Yu and Y.-H.~Zheng, \emph{{Operator bases
  in effective field theories with sterile neutrinos: d \ensuremath{\leq} 9}},
  \href{https://doi.org/10.1007/JHEP11(2021)003}{\emph{JHEP} {\bfseries 11}
  (2021) 003} [\href{https://arxiv.org/abs/2105.09329}{{\ttfamily
  2105.09329}}].

\bibitem{Pati:1974yy}
J.C.~Pati and A.~Salam, \emph{{Lepton Number as the Fourth Color}},
  \href{https://doi.org/10.1103/PhysRevD.10.275}{\emph{Phys. Rev. D} {\bfseries
  10} (1974) 275}.

\bibitem{Alioli:2017ces}
S.~Alioli, V.~Cirigliano, W.~Dekens, J.~de~Vries and E.~Mereghetti,
  \emph{{Right-handed charged currents in the era of the Large Hadron
  Collider}}, \href{https://doi.org/10.1007/JHEP05(2017)086}{\emph{JHEP}
  {\bfseries 05} (2017) 086}
  [\href{https://arxiv.org/abs/1703.04751}{{\ttfamily 1703.04751}}].

\bibitem{Abel:2020pzs}
C.~Abel et~al., \emph{{Measurement of the Permanent Electric Dipole Moment of
  the Neutron}},
  \href{https://doi.org/10.1103/PhysRevLett.124.081803}{\emph{Phys. Rev. Lett.}
  {\bfseries 124} (2020) 081803}
  [\href{https://arxiv.org/abs/2001.11966}{{\ttfamily 2001.11966}}].

\bibitem{Dekens:2018bci}
W.~Dekens, J.~de~Vries, M.~Jung and K.K.~Vos, \emph{{The phenomenology of
  electric dipole moments in models of scalar leptoquarks}},
  \href{https://doi.org/10.1007/JHEP01(2019)069}{\emph{JHEP} {\bfseries 01}
  (2019) 069} [\href{https://arxiv.org/abs/1809.09114}{{\ttfamily
  1809.09114}}].

\bibitem{Dekens:2018pbu}
W.~Dekens, E.E.~Jenkins, A.V.~Manohar and P.~Stoffer, \emph{{Non-perturbative
  effects in $\mu \to e \gamma$}},
  \href{https://doi.org/10.1007/JHEP01(2019)088}{\emph{JHEP} {\bfseries 01}
  (2019) 088} [\href{https://arxiv.org/abs/1810.05675}{{\ttfamily
  1810.05675}}].

\bibitem{Aebischer:2021uvt}
J.~Aebischer, W.~Dekens, E.E.~Jenkins, A.V.~Manohar, D.~Sengupta and
  P.~Stoffer, \emph{{Effective field theory interpretation of lepton magnetic
  and electric dipole moments}},
  \href{https://doi.org/10.1007/JHEP07(2021)107}{\emph{JHEP} {\bfseries 07}
  (2021) 107} [\href{https://arxiv.org/abs/2102.08954}{{\ttfamily
  2102.08954}}].

\bibitem{ACME:2018yjb}
{\scshape ACME} collaboration, \emph{{Improved limit on the electric dipole
  moment of the electron}},
  \href{https://doi.org/10.1038/s41586-018-0599-8}{\emph{Nature} {\bfseries
  562} (2018) 355}.

\bibitem{Beda:2013mta}
A.G.~Beda, V.B.~Brudanin, V.G.~Egorov, D.V.~Medvedev, V.S.~Pogosov,
  E.A.~Shevchik et~al., \emph{{Gemma experiment: The results of neutrino
  magnetic moment search}},
  \href{https://doi.org/10.1134/S1547477113020027}{\emph{Phys. Part. Nucl.
  Lett.} {\bfseries 10} (2013) 139}.

\bibitem{Borexino:2017fbd}
{\scshape Borexino} collaboration, \emph{{Limiting neutrino magnetic moments
  with Borexino Phase-II solar neutrino data}},
  \href{https://doi.org/10.1103/PhysRevD.96.091103}{\emph{Phys. Rev. D}
  {\bfseries 96} (2017) 091103}
  [\href{https://arxiv.org/abs/1707.09355}{{\ttfamily 1707.09355}}].

\bibitem{ALEPH:2013dgf}
{\scshape ALEPH, DELPHI, L3, OPAL, LEP Electroweak} collaboration,
  \emph{{Electroweak Measurements in Electron-Positron Collisions at
  W-Boson-Pair Energies at LEP}},
  \href{https://doi.org/10.1016/j.physrep.2013.07.004}{\emph{Phys. Rept.}
  {\bfseries 532} (2013) 119}
  [\href{https://arxiv.org/abs/1302.3415}{{\ttfamily 1302.3415}}].

\bibitem{CMS:2022yjm}
{\scshape CMS} collaboration, \emph{{Search for new physics in the lepton plus
  missing transverse momentum final state in proton-proton collisions at
  $\sqrt{s}$ = 13 TeV}},  \href{https://arxiv.org/abs/2202.06075}{{\ttfamily
  2202.06075}}.

\bibitem{deBlas:2013qqa}
J.~de~Blas, M.~Chala and J.~Santiago, \emph{{Global Constraints on Lepton-Quark
  Contact Interactions}},
  \href{https://doi.org/10.1103/PhysRevD.88.095011}{\emph{Phys. Rev. D}
  {\bfseries 88} (2013) 095011}
  [\href{https://arxiv.org/abs/1307.5068}{{\ttfamily 1307.5068}}].

\bibitem{Crivellin:2021bkd}
A.~Crivellin, M.~Hoferichter, M.~Kirk, C.A.~Manzari and L.~Schnell,
  \emph{{First-generation new physics in simplified models: from low-energy
  parity violation to the LHC}},
  \href{https://doi.org/10.1007/JHEP10(2021)221}{\emph{JHEP} {\bfseries 10}
  (2021) 221} [\href{https://arxiv.org/abs/2107.13569}{{\ttfamily
  2107.13569}}].

\bibitem{Allwicher:2022gkm}
L.~Allwicher, D.A.~Faroughy, F.~Jaffredo, O.~Sumensari and F.~Wilsch,
  \emph{{Drell-Yan Tails Beyond the Standard Model}},
  \href{https://arxiv.org/abs/2207.10714}{{\ttfamily 2207.10714}}.

\bibitem{CMS:2018ncu}
{\scshape CMS} collaboration, \emph{{Search for pair production of
  first-generation scalar leptoquarks at $\sqrt{s} =$ 13 TeV}},
  \href{https://doi.org/10.1103/PhysRevD.99.052002}{\emph{Phys. Rev. D}
  {\bfseries 99} (2019) 052002}
  [\href{https://arxiv.org/abs/1811.01197}{{\ttfamily 1811.01197}}].

\bibitem{Crivellin:2021egp}
A.~Crivellin, D.~M\"uller and L.~Schnell, \emph{{Combined constraints on first
  generation leptoquarks}},
  \href{https://doi.org/10.1103/PhysRevD.103.115023}{\emph{Phys. Rev. D}
  {\bfseries 103} (2021) 115023}
  [\href{https://arxiv.org/abs/2104.06417}{{\ttfamily 2104.06417}}].

\bibitem{Allwicher:2022mcg}
L.~Allwicher, D.A.~Faroughy, F.~Jaffredo, O.~Sumensari and F.~Wilsch,
  \emph{{HighPT: A Tool for high-$p_T$ Drell-Yan Tails Beyond the Standard
  Model}},  \href{https://arxiv.org/abs/2207.10756}{{\ttfamily 2207.10756}}.

\bibitem{CMS:2021ctt}
{\scshape CMS} collaboration, \emph{{Search for resonant and nonresonant new
  phenomena in high-mass dilepton final states at $ \sqrt{s} $ = 13 TeV}},
  \href{https://doi.org/10.1007/JHEP07(2021)208}{\emph{JHEP} {\bfseries 07}
  (2021) 208} [\href{https://arxiv.org/abs/2103.02708}{{\ttfamily
  2103.02708}}].

\bibitem{ATLAS:2019lsy}
{\scshape ATLAS} collaboration, \emph{{Search for a heavy charged boson in
  events with a charged lepton and missing transverse momentum from $pp$
  collisions at $\sqrt{s} = 13$ TeV with the ATLAS detector}},
  \href{https://doi.org/10.1103/PhysRevD.100.052013}{\emph{Phys. Rev. D}
  {\bfseries 100} (2019) 052013}
  [\href{https://arxiv.org/abs/1906.05609}{{\ttfamily 1906.05609}}].

\bibitem{Chen:1969hkp}
H.H.~Chen, \emph{{Electromagnetic simulation of time-reversal violation in
  mirror spin 3/2 beta decays}},
  \href{https://doi.org/10.1103/PhysRev.185.2003}{\emph{Phys. Rev.} {\bfseries
  185} (1969) 2003}.

\bibitem{Holstein:1972bbl}
B.R.~Holstein, \emph{{Tests for t invariance in allowed nuclear beta decay}},
  \href{https://doi.org/10.1103/PhysRevC.5.1529}{\emph{Phys. Rev. C} {\bfseries
  5} (1972) 1529}.

\bibitem{Ando:2009jk}
S.-i.~Ando, J.A.~McGovern and T.~Sato, \emph{{The D coefficient in neutron beta
  decay in effective field theory}},
  \href{https://doi.org/10.1016/j.physletb.2009.04.088}{\emph{Phys. Lett. B}
  {\bfseries 677} (2009) 109}
  [\href{https://arxiv.org/abs/0902.1194}{{\ttfamily 0902.1194}}].

\bibitem{Gonzalez-Alonso:2017iyc}
M.~Gonz\'alez-Alonso, J.~Martin~Camalich and K.~Mimouni,
  \emph{{Renormalization-group evolution of new physics contributions to
  (semi)leptonic meson decays}},
  \href{https://doi.org/10.1016/j.physletb.2017.07.003}{\emph{Phys. Lett. B}
  {\bfseries 772} (2017) 777}
  [\href{https://arxiv.org/abs/1706.00410}{{\ttfamily 1706.00410}}].

\bibitem{Cirigliano:2021yto}
V.~Cirigliano, D.~D\'\i{}az-Calder\'on, A.~Falkowski, M.~Gonz\'alez-Alonso and
  A.~Rodr\'\i{}guez-S\'anchez, \emph{{Semileptonic tau decays beyond the
  Standard Model}},  \href{https://arxiv.org/abs/2112.02087}{{\ttfamily
  2112.02087}}.

\bibitem{Buchmuller:1986zs}
W.~Buchmuller, R.~Ruckl and D.~Wyler, \emph{{Leptoquarks in Lepton - Quark
  Collisions}}, \href{https://doi.org/10.1016/0370-2693(87)90637-X}{\emph{Phys.
  Lett. B} {\bfseries 191} (1987) 442}.

\bibitem{Dorsner:2016wpm}
I.~Dor\v{s}ner, S.~Fajfer, A.~Greljo, J.F.~Kamenik and N.~Ko\v{s}nik,
  \emph{{Physics of leptoquarks in precision experiments and at particle
  colliders}}, \href{https://doi.org/10.1016/j.physrep.2016.06.001}{\emph{Phys.
  Rept.} {\bfseries 641} (2016) 1}
  [\href{https://arxiv.org/abs/1603.04993}{{\ttfamily 1603.04993}}].

\bibitem{Dorsner:2019itg}
I.~Dor\v{s}ner, S.~Fajfer and O.~Sumensari, \emph{{Muon $g-2$ and scalar
  leptoquark mixing}},
  \href{https://doi.org/10.1007/JHEP06(2020)089}{\emph{JHEP} {\bfseries 06}
  (2020) 089} [\href{https://arxiv.org/abs/1910.03877}{{\ttfamily
  1910.03877}}].

\end{thebibliography}\endgroup

\end{document}